\documentclass[journal]{IEEEtran}
\usepackage{amsmath,epsfig}
\usepackage{color}
\usepackage{xcolor}
\usepackage{amsthm,amssymb,bm}
\usepackage{comment}
\usepackage{subcaption}
\usepackage{soul}
\usepackage{hyperref}
\usepackage{cite}

\newcommand{\done}{\rlap{$\square$}\raisebox{.15ex}{\hspace{0.1em}$\checkmark$}
\hspace{-1pt}}

\newcommand\huy{\textcolor{black}}

\newcommand\Qunit{$\mathrm{m}^{3}.\mathrm{s}^{-1}$}
\newcommand\Ksunit{$\mathrm{m}^{1/3}.\mathrm{s}^{-1}$}

\title{Assimilation of SWOT Altimetry Data for Riverine Flood Reanalysis: From Synthetic to Real Data}
\author{Quentin~Bonassies, 
        Thanh~Huy~Nguyen,~\IEEEmembership{Member,~IEEE,}
        Ludovic~Cassan,
        Andrea~Piacentini,
        Sophie~Ricci,
        Charlotte~Emery,
        Christophe~Fatras,
        Santiago~Pe\~{n}a~Luque,
        and Raquel~Rodriguez~Suquet
\thanks{This work was supported in part by the Centre National d'\'{E}tudes Spatiales (CNES) and in part by the Centre Europ\'{e}en de Recherche et de Formation Avanc\'{e}e en Calcul Scientifique (CERFACS). This research was funded by the Luxembourg National Research Fund (FNR) through SWIFT (Grant no. INTER/ANR/23/17800438) project, and by the Agence Nationale de la recherche (France): ANR-23-CE56-0009.
(Corresponding author: Thanh Huy Nguyen.)}
\thanks{Quentin Bonassies, Ludovic Cassan and Sophie Ricci are with the CERFACS, 31057 Toulouse Cedex 1, France, and also with the CECI Laboratory, CERFACS/CNRS UMR 5318, 31057 Toulouse Cedex 1, France (e-mail: bonassies@cerfacs.fr, cassan@cerfacs.fr, ricci@cerfacs.fr).}
\thanks{Thanh Huy Nguyen was with the CERFACS, 31057 Toulouse Cedex 1, France, and also with the CECI Laboratory, CERFACS/CNRS UMR 5318, 31057 Toulouse Cedex 1, France. He is now with the Environmental Research and Innovation department, Luxembourg Institute of Science and Technology (LIST), L-4326 Esch-sur-Alzette, Luxembourg  (e-mail: thanh-huy.nguyen@list.lu).}
\thanks{Andrea Piacentini is with the CERFACS, 31057 Toulouse Cedex 1, France
(piacentini.palm@gmail.com).}
\thanks{Charlotte~Emery is with CS-Group, 31401 Toulouse Cedex 9, France
(charlotte.emery@cs-soprasteria.com).}
\thanks{Christophe~Fatras is with Collecte Localisation Satellites (CLS), Ramonville Saint-Agne, 31520, France
(cfatras@groupcls.com).}
\thanks{Santiago Pe\~{n}a Luque and Raquel Rodriquez Suquet are with the Centre National d'\'{E}tudes Spatiales (CNES), 31401 Toulouse Cedex 9, France (e-mail: santiago.penaluque@cnes.fr, raquel.rodriguezsuquet@cnes.fr).}
\thanks{Manuscript received April 19, 2025; revised August 16, YYYY.}
}

\begin{document}
%
\maketitle
\begin{abstract}
Floods are one of the most common and devastating natural disasters worldwide. 
The contribution of remote sensing is important for reducing the impact of flooding both during the event itself and for improving hydrodynamic models by reducing their associated uncertainties. 
%
This article presents the innovative capabilities of the Surface Water and Ocean Topography (SWOT) mission, especially its river node products, to enhance the accuracy of riverine flood reanalysis, 
performed on a 50-km stretch of the Garonne River.
The experiments incorporate various data assimilation strategies, based on the ensemble Kalman filter (EnKF), which allows for sequential updates of model parameters based on available observations. The experimental results show that while SWOT data alone offers some improvements, combining it with in-situ water level measurements provides the most accurate representation of flood dynamics, both at gauge stations and along the river. The study also investigates the impact of different SWOT revisit frequencies on the model's performance, revealing that assimilating more frequent SWOT observations leads to more reliable flood reanalyses. In the real event, it was demonstrated that the assimilation of SWOT and in-situ data accurately reproduces the water level dynamics, offering promising prospects for future flood monitoring systems. Overall, this study emphasizes the complementary strengths of Earth Observation data in improving the representation of the flood dynamics in the riverbed and the floodplains.
\end{abstract}


\section{Introduction}
\label{sec:intro}

\huy{Future flood risk is anticipated to rise due to urban growth and unchecked urbanization of floodplains, which collectively contribute to heightened exposure and vulnerability to flooding events \cite{doi/10.2760/14505,alfieri2016increasing,smith2019new}.}
\huy{In this context, the integration of satellite-based Earth Observation (EO) data offers substantial potential for improving flood risk forecasting models. Enhanced model performance, supported by EO data, contributes to more effective flood mitigation strategies and the protection of vulnerable infrastructure and assets. In particular, remote sensing (RS) observations of water surface dynamics during flood events provide insights for in-depth hydrodynamic analysis and investigation \cite{Schumann2009,Bates2004}.}
\huy{Though RS data provide valuable spatially distributed observations, their relatively low revisit frequency limits their utility for resolving the rapid temporal dynamics of flood events. Recent efforts have addressed this limitation through the integration of observations from multiple satellite missions \cite{boergens2017combination,tourian2016spatiotemporal,zakharova2020river,nguyen2024remote,la2024early}. In contrast, in-situ sensors offer high temporal resolution but are spatially sparse, providing data at only a limited number of  locations along river networks. Consequently, neither data source alone is sufficient to fully characterize the spatiotemporal complexity of flood dynamics. Combining them offers a promising path forward, enabling improved representation of flood dynamics through complementary strengths.
}

The TELEMAC-2D \cite{hervouet2007hydrodynamics} code solves depth-averaged free surface flow  Navier-Stokes equations, known as Shallow water equations as derived first by Barré de Saint Venant \cite{de1871theorie}. 
TELEMAC-2D offers significant advantages in river hydraulics and flood simulation by providing high-fidelity hydrodynamic modeling, while its parallel computing capabilities are suitable for large-scale applications \cite{moulinec2011telemac}, especially in ensemble data assimilation (DA).
When being used as a forward model in a DA framework, the model outputs are compared to available observations such as water level (WL) measurements and/or flood extent observations \cite{Madsen2005, Neal2007, Neal2009, leedal2010visualization} to allow for the sequentially updating and correcting parameters and inputs. 
DA strategies can employ either variational methods or ensemble Kalman filters (EnKF) \cite{evensen2003ensemble}. The EnKF methods, which rely on computing stochastically the forecast error covariance matrices through a limited number of perturbed simulations, can be adapted for the context of free-surface flow \cite{NguyenTGRS2022, nguyenagu2022, nguyen2023gaussian}, resulting in more accurately simulation of real-world flood events.

Similar to in-situ WLs, the assimilation of RS-derived hydrometric WLs significantly is straightforward, especially when compared to the assimilation of flood extents \cite{hostache2018near}. 
However, since RS-derived WLs are conventionally obtained by combining the observed flood extents with a digital elevation model (DEM) \cite{giustarini2011assimilating,yang2024remote}, their retrieval accuracy strongly depends on the quality and resolution of the DEM. Furthermore, because the DEM is used both as an input for the hydraulic model and for deriving observed water levels, an inherent bias arises due to the lack of independence between the observations and model inputs \cite{schumann2008comparison}.
A review by Grimaldi \textit{et al.} \cite{grimaldi2016remote} provides a comprehensive understanding of the value of coarse, medium, and high resolution RS observations for flood extent and WL monitoring. Additionally, the review emphasizes the importance of incorporating in-situ data, particularly when hydrodynamic models are utilized. 
In this regard, altimetry data are highly beneficial for flood reanalysis and forecasting. 
Water Surface Elevation (WSE) is measured from space by several nadir satellite altimetry missions, including TOPEX/Poseidon, Jason-1/-2/-3, SARAL/AltiKa, Sentinel-3, and Sentinel-6. 

Among satellite altimetry missions, the Surface Water and Ocean Topography (SWOT) wide-swath altimetry satellite represents a significant advancement in oceanography and hydrology \cite{fu2024surface}, even long before its launch in December 2022. It offers global coverage of Water Surface Elevations (WSE) at high resolution with a 21-day revisit frequency, covering a 120~km-wide swath (with a roughly 20~km gap at nadir). 
Numerous studies were conducted in preparation for such a mission \cite{Emery2020, brisset:hal-02044488, OUBANAS2018638}, including the development of data simulators to create algorithms for continental water processing \cite{emery2022tools4swotsims}. 
The merits of SWOT data for the calibration of channel and floodplain roughness, as well as for the estimation of river discharge were demonstrated in~\cite{wongchuig2020assimilation,de2021using,durand2023framework,oubanas2024hydraulic}. 
New research opportunities for improving flood forecasting lie in the water depth maps and/or flood extent maps derived from SWOT images. 
Although the quality of WSE time-series data can be affected by many factors, including land cover, floodplain topography and channel morphology \cite{de2021using,maillard2015new}, the extensive applications of satellite altimetry underscore its significant potential across numerous environments.

\huy{
This study investigates the potential of assimilating SWOT altimetry data for riverine flood applications, progressing from controlled synthetic experiments to the assimilation of real satellite observations. The objective is to assess the merits of SWOT river products to enhance flood reanalysis using the TELEMAC-2D hydrodynamic model. To this end, an Observing System Simulation Experiment (OSSE) is first conducted, employing synthetic EO data generated from a reference hydraulic simulation. This framework serves as a foundation for the subsequent real-data application, allowing for controlled assessment of how SWOT assimilation reduces uncertainties in model inputs and parameters.
The OSSE framework also provides the flexibility to test different SWOT revisit frequencies, thereby evaluating their impact on reanalysis skill and highlighting the benefits of more frequent satellite observations for flood forecasting. In the real-data assimilation experiment, the focus shifts to identifying methodological adjustments needed to effectively incorporate actual SWOT measurements. Particular attention are dedicated to analyzing the discrepancies between the OSSE and real recent event, which can stem from observational errors, unaccounted in the synthetical data generation. Although the 2024 event did not result in overbank flooding in the study area, meaningful insights were still gained through analysis of in-channel flow dynamics, demonstrating the relevance of SWOT data even under non-flooding conditions.
}

\section{Material and Method}
\label{sec:method}

\subsection{Study area}
\label{subsect:TELEMAC-2DGaronne}

\begin{figure*}[t]
\centering
    \begin{subfigure}[b]{0.48\linewidth}
    \includegraphics[width=\linewidth]{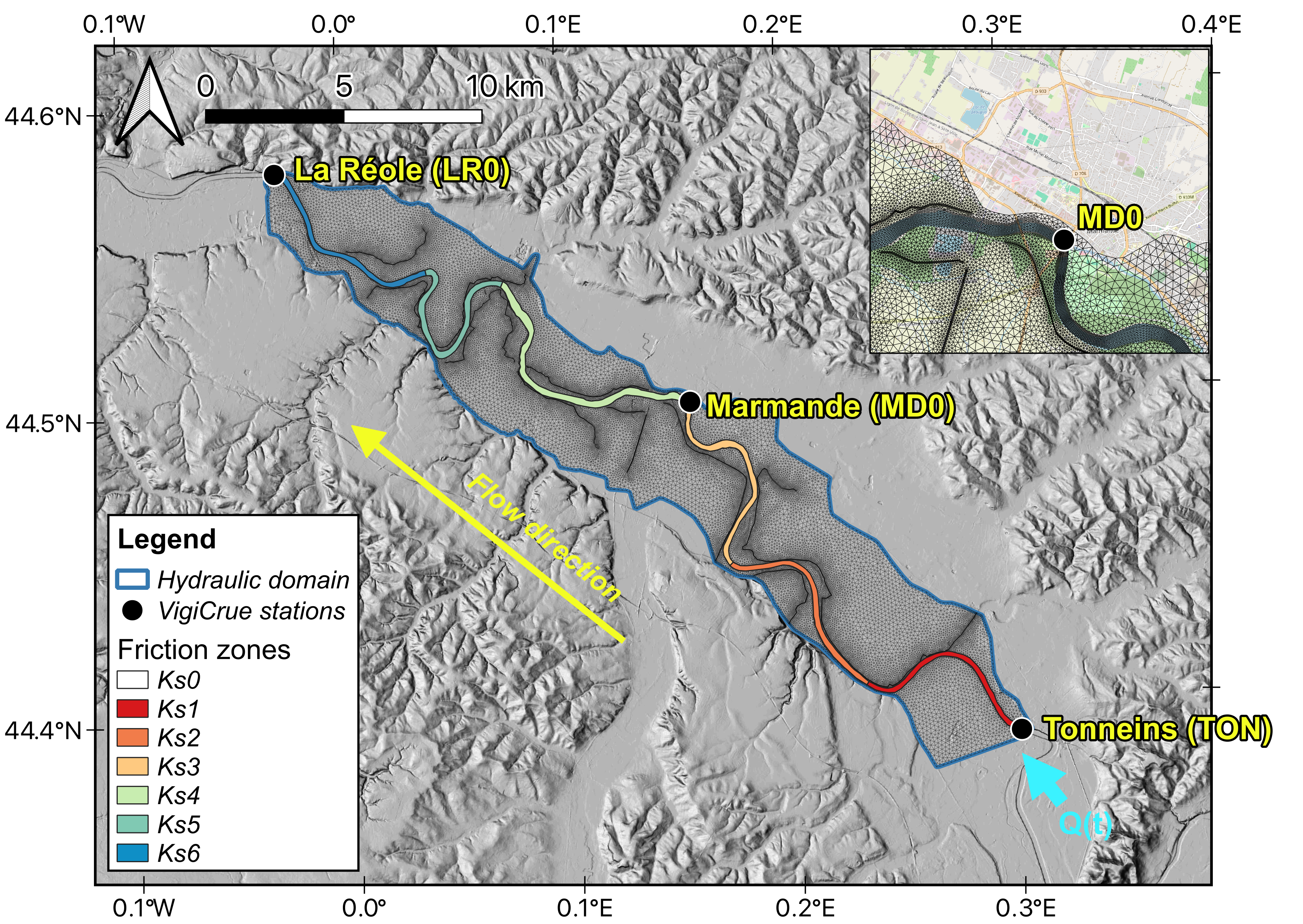}
    \caption{OSSE}
    \end{subfigure}
    \begin{subfigure}[b]{0.48\linewidth}
    \includegraphics[width=\linewidth]{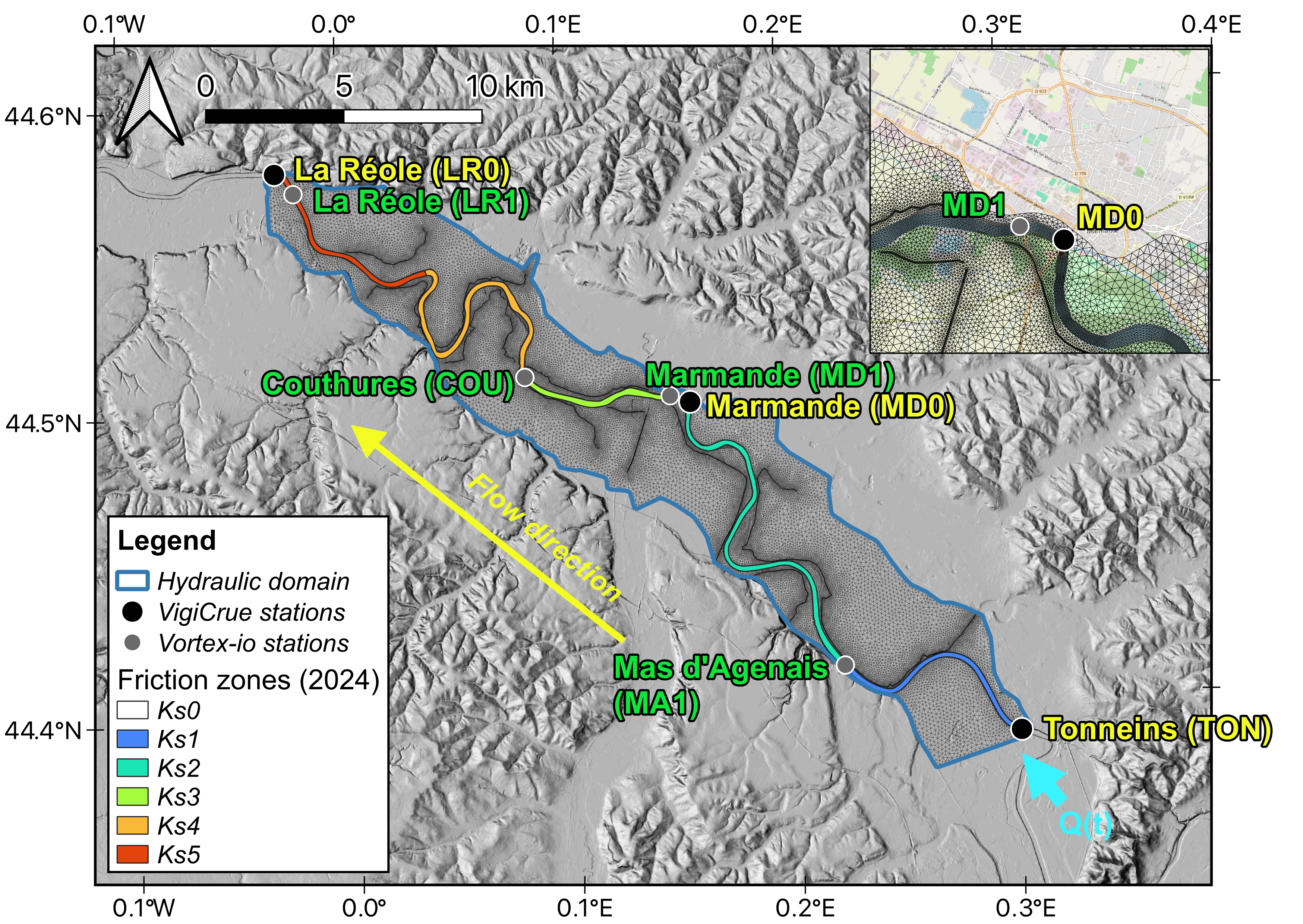}
    \caption{2024 real event}
    \end{subfigure}
     \caption{The TELEMAC-2D Garonne Marmandaise domain (within blue outline). The gauge stations are indicated as black (VigiCrue) and gray (Vortex-io) circles. The different river friction zones are indicated as colored segments of the riverbed. The inset figure at the top-right corner magnifies the urban area of Marmande and thereby gauge stations. The left panel is the configuration for OSSE (with six riverbed friction zones) and the right panel is for real event (with five riverbed friction zones). Basemap: \copyright~ESRI World Hillshade \cite{ESRI}.}
     \label{fig:study_area}
\end{figure*}

The study area covers a 50-km stretch of the Garonne River in southwest France, extending from Tonneins, downstream of its confluence with the Lot River, to the commune of La Réole (Figure~\ref{fig:study_area}), historically impacted by floods. This region has a strong agricultural component. Since the 19th century, various infrastructures have been developed to safeguard the Garonne floodplain from major flooding events, such as the historic 1875 flood. Over time, a network of longitudinal dikes and weirs has been gradually built to protect and cultivate floodplains \cite{Marchandise31122024}.

\begin{figure}[!t]
\centering
    \includegraphics[width=\linewidth]{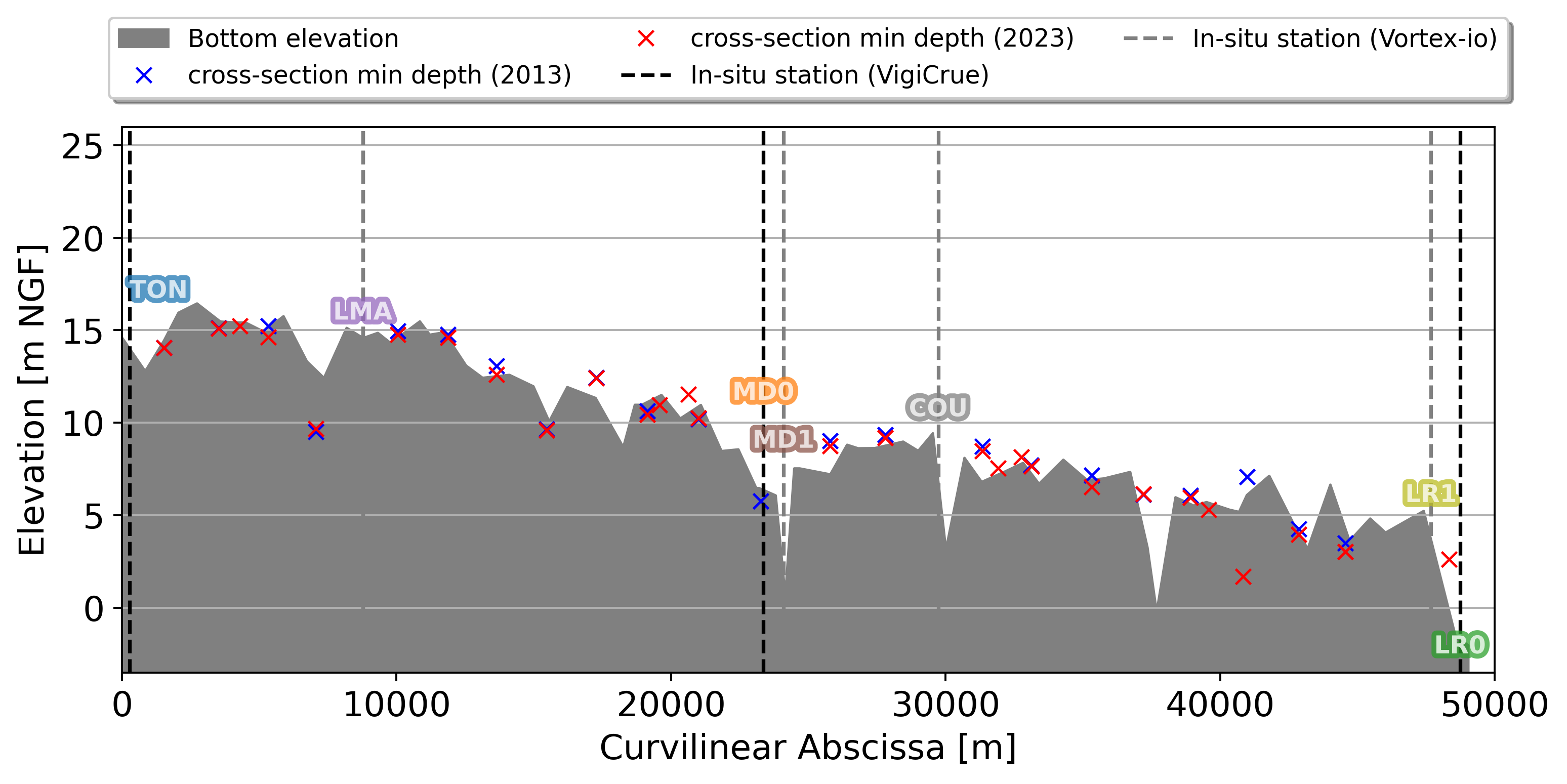}
     \caption{Bottom elevation profile of the study domain. Blue line is extracted on the centerline from TELEMAC-2D model. Vertical lines represent the location of gauge stations, 
     whereas red and blue crosses represent the bathymetric surveys.}
     \label{fig:profile}
     
\end{figure}

A hydrodynamic model with a 41,000-node mesh is built with the TELEMAC-2D \cite{hervouet2007hydrodynamics} from bathymetric cross-sectional profiles (Figure~\ref{fig:profile}) and topographic data. 
Boundary conditions (BCs) are provided by a downstream rating curve (La Réole gauge station) and an upstream hydrograph (Tonneins gauge station). It should be noted that the inflow discharge at Tonneins ($Q_{BC}$) is one of the control variables in the assimilation process. It is corrected by a multiplying factor $\mu$ to the upstream hydrograph. The friction coefficient is defined over several zones, including several segments $K_s$. 
A detailed description of the TELEMAC-2D hydrodynamic model is given in \cite{nguyenagu2022}. 

Observing stations from the national gauge network VigiCrue, are located at Tonneins, Marmande and La Réole, respectively abbreviated as TON, MD0 and LR0.
Their hydrometric data can be accessed from HydroPortail\footnote{\url{https://hydro.eaufrance.fr/}}, a portal of data from the Flood Forecasting Services (SPC) measurement network. \huy{The initial calibration of the TELEMAC-2D model was performed using in-situ observations from the three VigiCrue stations, based on a series of non-overflowing events \cite{besnard2011comparaison}.}
Existing works on the same catchment \cite{nguyenagu2022,nguyen2023gaussian} proposed an enhanced setting of the TELEMAC-2D model, where the river channel is divided into six friction zones, using Strickler coefficients \cite{gauckler1867etudes} denoted by $K_{s_i}$ (with $i \in [1,..,6]$), as depicted by the solid colored segments in Figure~\ref{fig:study_area}. However, this setup has a limitation related to the number of available in-situ gauge stations, with some zones lacking direct measurements during past flood events, such as $K_{s_2}$ and $K_{s_5}$. 

The SCO-FloodDAM project, initiated in 2021 \cite{kettig,10282907}, in collaboration with NASA \cite{2023AGUFMIN51A..03H}, deployed additional gauge micro-stations \cite{gachelin2025development} across the study area, which were operational during the 2024 event. Stations at Le Mas-d'Agenais, Marmande, Couthures-sur-Garonne, and La Réole, provided by Vortex-io\footnote{\url{https://www.vortex-io.fr/}}, contributed to increasing the total number of observing stations to seven, as summarized in Table~\ref{tab:in-situ} in the order of increasing curvilinear abscissa along the study river reach. Benefiting from the improved observation network density and building on the experience from previous studies \cite{nguyenagu2022,nguyen2023gaussian}, the number of Strickler zones for the 2024 event was reduced from six to five, as shown in Figure~\ref{fig:study_area}. Consequently, the TELEMAC-2D model configuration for the 2024 event comprises five Strickler zones, each associated with at least one gauge station.

Figure~\ref{fig:profile} depicts the riverbed elevation along the river centerline extracted from the TELEMAC-2D mesh. In order to better analyze results, the river bottom represented by cross-sections, acquired from two  bathymetric surveys in 2013 and 2023, has also been shown here. These cross-sections are provided by the SPC Garonne-Tarn-Lot \cite{Marchandise31122024}. They are shown by red and blue crosses in Figure~\ref{fig:profile}. It reveals that the river bathymetry has changed in some areas where results would differ from simulations.

\begin{table*}[h]
    \centering
    \caption{In-situ data at VigiCrue and Vortex-io observing stations used for the 2024 flood event. The stations are listed in order of increasing river kilometer along the study reach. 
    }
    \label{tab:in-situ}
    \begin{tabular}{cc|ccccc|cc}
    \hline
    & & Network/ & Installation & Physical & & Time & \multicolumn{2}{|c}{Strickler zone} \\
    Stations & Abbre. & Provider & date & location & Utilization &  interval & OSSE & Real event \\ \hline
    Tonneins & TON & VigiCrue & 1989-01-01 & Tonneins bridge & Assimilation & 15 mins & $K_{s_1}$ & $K_{s_1}$ \\
    Le Mas-d'Agenais & LMA & Vortex-io & 2022-11-26 & Le Mas-d'Agenais bridge & Assimilation & 60 mins & $K_{s_2}$ & $K_{s_2}$ \\
    Marmande & MD0 & VigiCrue & 1986-03-14 & Renaud Jean bridge & Assimilation & 15 mins & $K_{s_4}$ & $K_{s_3}$ \\
    \textit{Marmande 1} & \textit{MD1} & Vortex-io & 2022-02-17 &  D933 road bridge & \textit{Validation} & 60 mins & $K_{s_4}$ & $K_{s_3}$\\
    Couthures-sur-Garonne & COU & Vortex-io & 2021-02-18 & D3 road bridge & Assimilation & 60 mins & $K_{s_5}$ & $K_{s_4}$ \\
    \textit{La Réole 1} & \textit{LR1} & Vortex-io & 2022-04-08 & D9 road bridge & \textit{Validation} & 60 mins & $K_{s_6}$ & $K_{s_5}$ \\
    La Réole & LR0 & VigiCrue & 1990-01-01 & Rouergue bridge & Assimilation & 15 mins & $K_{s_6}$ & $K_{s_5}$ \\
    \hline
    \end{tabular}
\end{table*}

\subsection{Remote sensing data}

The SWOT satellite is a pioneering mission designed to provide high-resolution measurements of Earth's surface water. Launched as a collaboration between NASA and CNES, it leverages advanced altimetry technology to map and monitor global surface WLs with unprecedented accuracy. Specifically for inland waters and river monitoring, SWOT  can track the height, extent, and dynamics of lakes, rivers, reservoirs, and floodplains \cite{fu2024surface}. It offers valuable insights into hydrological processes, water management, and climate change impacts by delivering near-real-time information on the world's freshwater resources \cite{de2021using,durand2023framework}. 


\huy{Sentinel-6 is a Copernicus mission dedicated primarily to monitoring Earth's oceans with exceptional accuracy. Its aims at enhancing the understanding of climate change impacts by providing precise measurements related to sea-level rise, ocean circulation, and climate variability. Additionally, Sentinel-6 delivers high-quality observations relevant to continental hydrology.
In this study, we analyze data obtained from Sentinel-6’s Poseidon-4 Ku/C-band nadir-pointing SAR altimeter. Poseidon-4 operates in high-resolution mode, employing an interleaved chronogram acquisition \cite{DINARDO2024337}. Applying the Fully-Focused SAR (FFSAR) processing technique \cite{egido2016fully} to these measurements enables unprecedented accuracy in nadir altimetry for inland water bodies, achieved through fully coherent processing of radar pulse echoes from the pulse-limited, nadir-looking altimeter.}
Yet, in order to maintain a robust signal-to-noise-ratio exceeding 20~dB, along-track distance from the satellite track to the river cannot be greater than 2~km. Under these conditions, off-nadir signals enable the estimation of longitudinal profiles of WSEs \cite{boy2023measuring}. 
\huy{As such, this technique demonstrates applicability over an 18-km stretch of the Garonne River within the study area. Sentinel-6 altimetry data were processed using the FFSAR technique \cite{egido2016fully}, providing spatially dense WSE profiles sampled every 10~m along the river centerline at each satellite overpass. Sentinel-6 altimetry data processed using the FFSAR technique have also demonstrated strong potential for monitoring terrestrial water bodies and floods \cite{2023EGUGA..25..449M,2023EGUGA..25..448G,2023EGUGA..25.6513R,nguyen2024chained}, exhibiting reliability even under challenging conditions, such as sandy riverbanks or the presence of hydraulic structures (e.g., locks and dams) \cite{boy2023measuring}. However, the narrow swath coverage of Sentinel-6 remains a limitation, highlighting the necessity to integrate these data with additional EO datasets.}

In this study, SWOT data serve as the primary source, in combination with in-situ data, for DA in the experiments conducted on both the OSSE and the real event. 
The studied 50-km stretch of the Garonne River is covered by three passes 42, 113 and 391, as illustrated by Figure~\ref{fig:swot_passplan}. 
\huy{While it is evident that passes 42 and 113 only partially cover the hydraulic domain in reality, their corresponding SWOT observations are retained with full spatial coverage (like that of pass 391) in the OSSE setup to maximize the potential benefits of SWOT data for assimilation purposes.}
Conversely, Sentinel-6 FFSAR measurements are used for validation through longitudinal profiles along the river, while in-situ data provide point observations at the gauge stations. 

\subsection{Observing System Simulation Experiment}
\label{ssec:osse}
\begin{figure}[!t]
\centering
    \includegraphics[width=\linewidth]{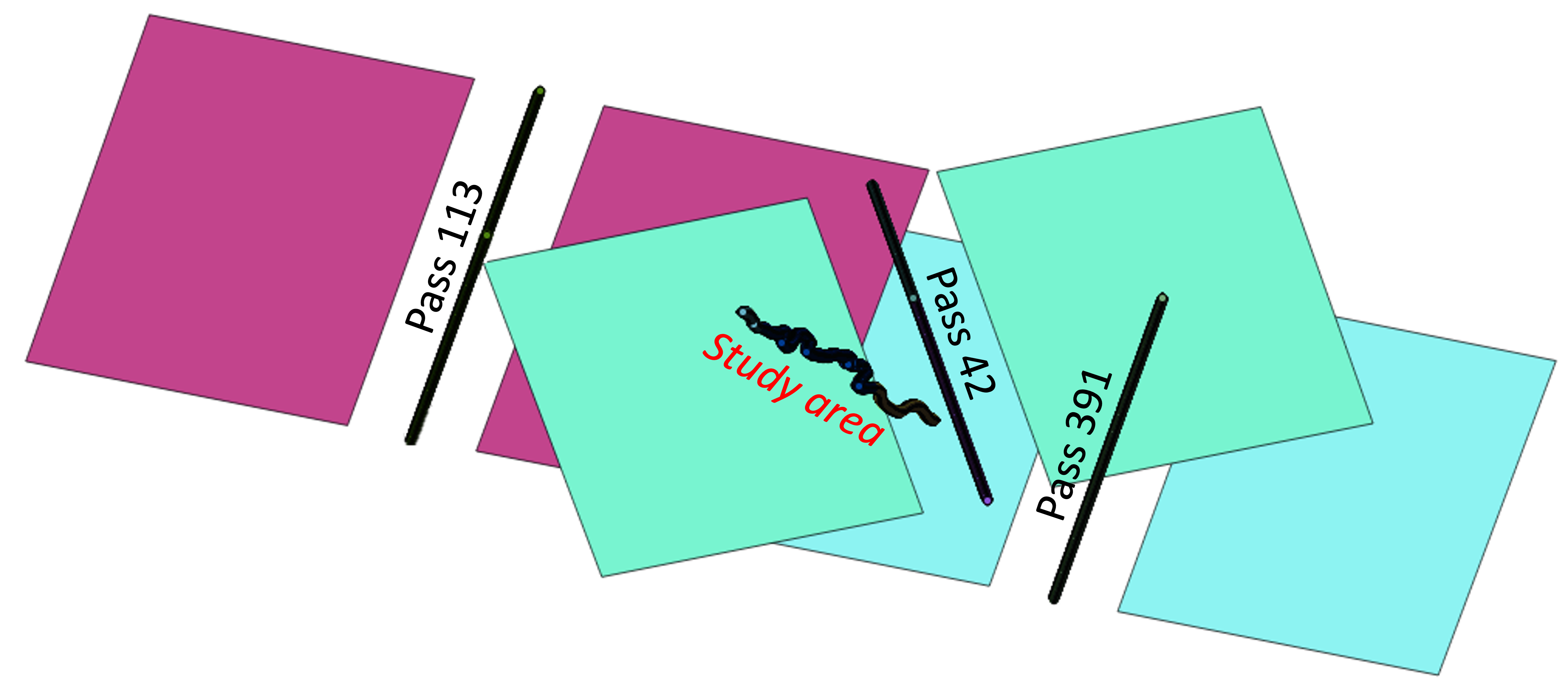}
     \caption{SWOT passes over the study area.}
     \label{fig:swot_passplan}
\end{figure}

\begin{figure}[!t]
\begin{subfigure}[b]{\linewidth}
\centering
\includegraphics[trim=0 0 0 0, clip, width=\linewidth]{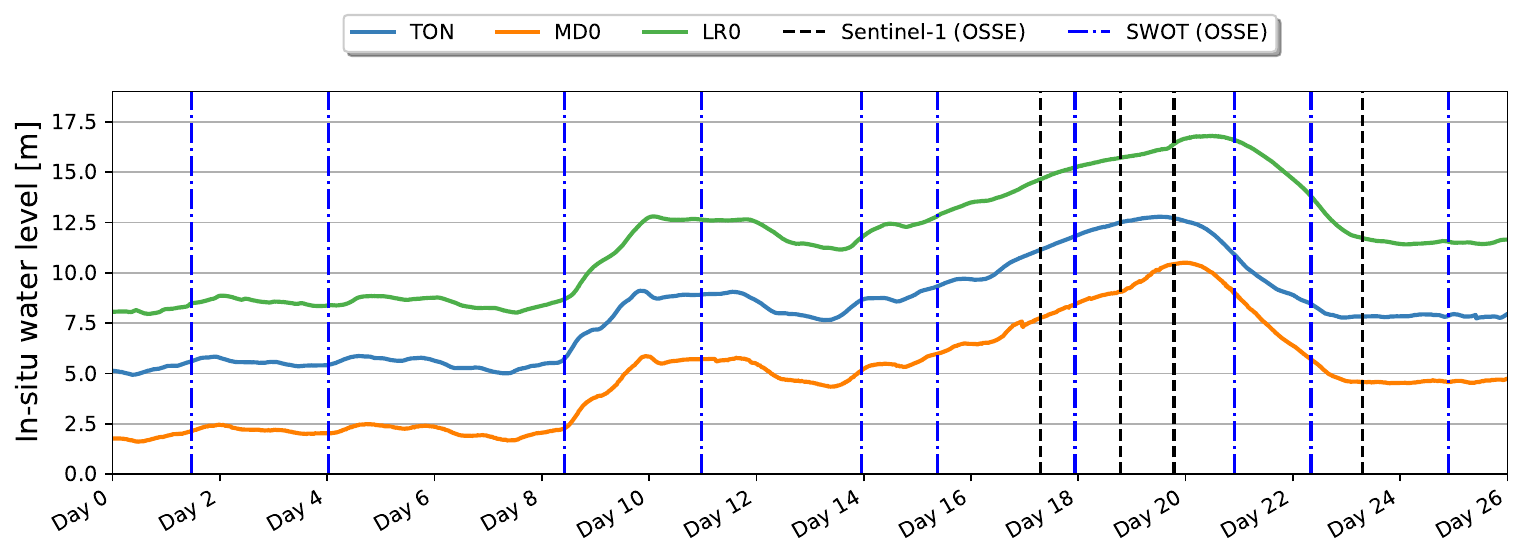}
\caption{OSSE}\label{fig:obs_OSSE}
\end{subfigure}

\begin{subfigure}[b]{\linewidth}
\centering
\includegraphics[trim=0 0 0 0, clip, width=\linewidth]{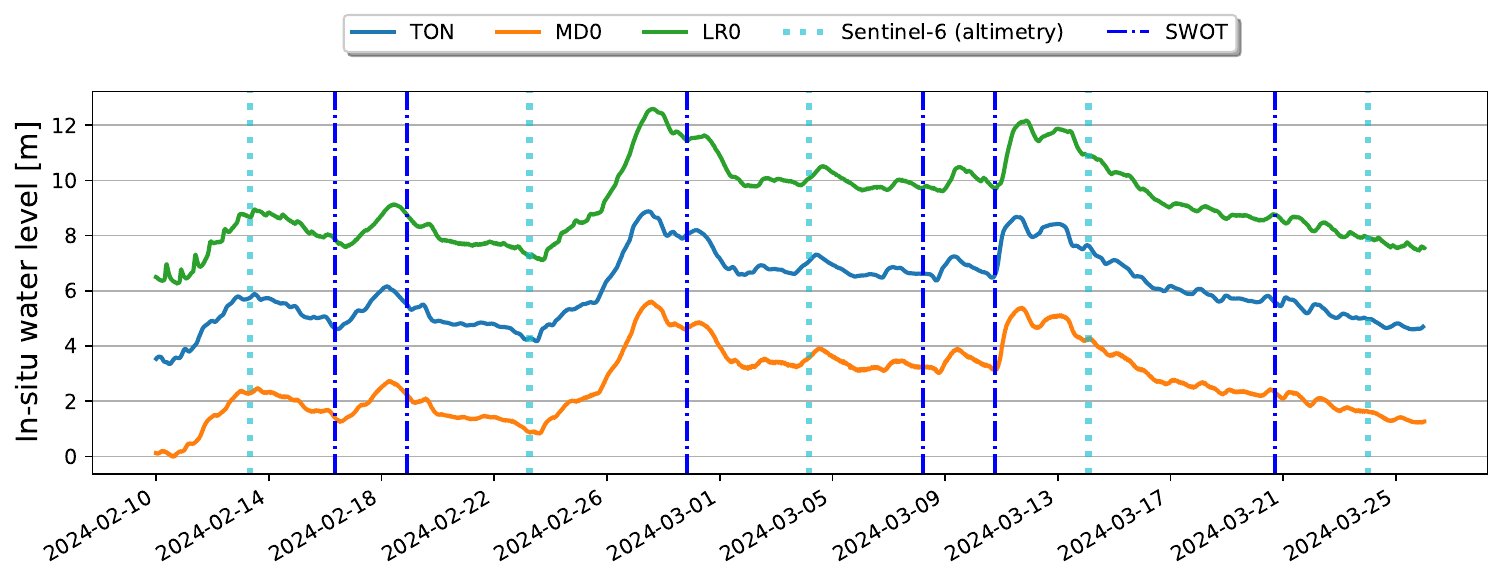}
\caption{2024 real event}\label{fig:obs_2024}
\end{subfigure}

\caption{In-situ WL time-series at VigiCrue stations: Tonneins (blue), Marmande (orange) and La Réole (green) during the (a) synthetical event and (b) 2024 real event.}
\label{fig:obs}
\end{figure}

The OSSE observation setting was based on a real flood scenario that occurred in 2021 \cite{nguyenagu2022}. Here, the SWOT pass plan (Figure~\ref{fig:swot_passplan}) was synthetically enhanced, compared to the 21-day repeat cycle of the nominal science orbit, by tripling the number of passes during the flood period \cite{nguyen2024assimilation}. This allows a temporal sampling of 2-3 days for SWOT over the studied catchment, which is more ideal for flood studies. Figure~\ref{fig:obs} depicts the in-situ WLs for the synthetical flood event at VigiCrue observing stations, namely Tonneins (TON), Marmande (MD0) and La Réole (LR0), in blue, orange and green lines, respectively. 

\begin{figure}[!t]
\centering
\begin{subfigure}[b]{\linewidth}
\centering
\includegraphics[trim=0 0cm 0 0,clip,width=\linewidth]{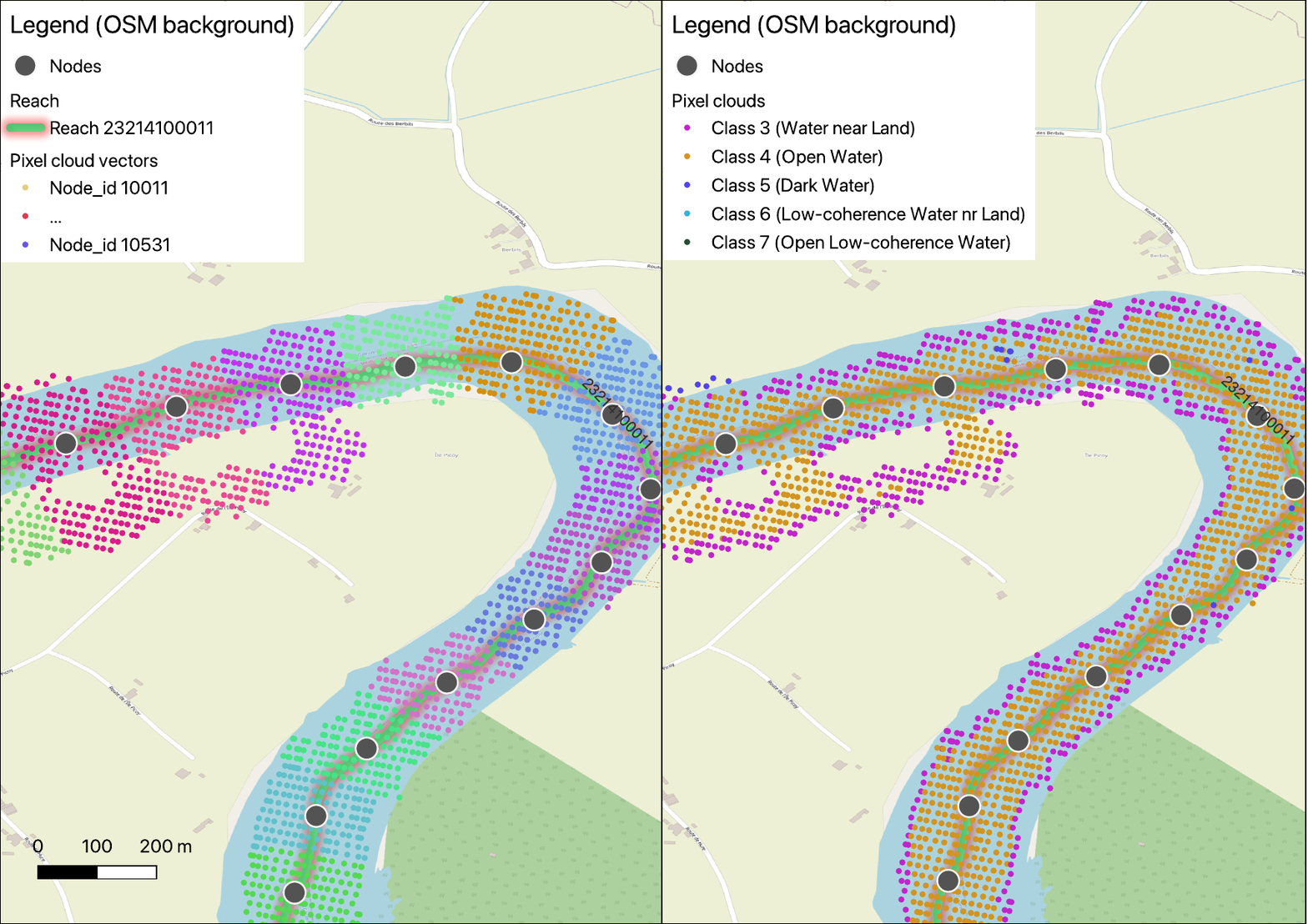}
\caption{OSSE}\label{fig:SWOT_nodes_OSSE}
\end{subfigure}\vspace{0.25cm}

\begin{subfigure}[b]{\linewidth}
\centering
\includegraphics[trim=0 0cm 0 0,clip,width=\linewidth]{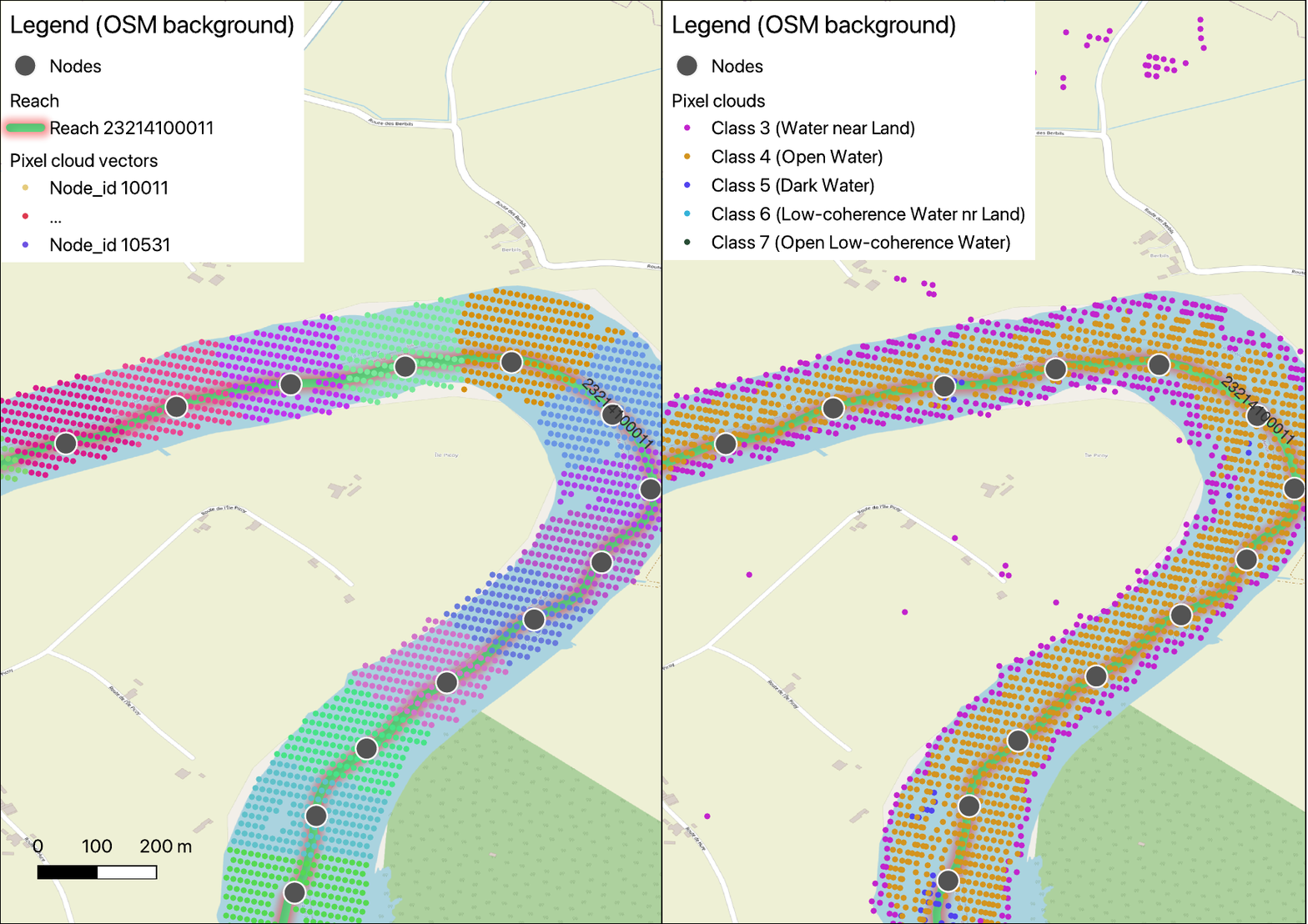}
\caption{2024 real event}\label{fig:SWOT_nodes_real}
\end{subfigure}
\caption{Node-level WSE SWOT river products: (a) synthetical SWOT data on day 8 (OSSE) and (b) real SWOT data on 2024-02-28 (real event). Left panels: color-coded aggregated pixels for each node; right panels: classification of pixels. Basemap: OpenStreetMap (obtained through QuickMapServices QGIS plugin).}
\label{fig:SWOT_nodes}
\end{figure}

The synthetical SWOT river products are generated from the reference water depth maps at SWOT overpass times following the configured pass plan. The reference is extracted from reanalysis of the real 2021 flood event \cite{nguyenagu2022}. Parameters and water depth in the floodplain have been smoothed to represent a more realistic flood dynamics. The maps are first processed using the Tools4SWOTsims toolbox \cite{emery2022tools4swotsims}. This toolbox consists of a set of Python scripts to map 1D/2D hydrodynamic model outputs into 2D WSE rasters that are compatible with the SWOT-HR simulator\footnote{\texttt{https://github.com/CNES/swot-hydrology-toolbox}}, dedicated to hydrological science \cite{frasson2017}. The output of the SWOT-HR simulator is SWOT observations in pixel cloud format. Such pixel cloud data are then processed by the RiverObs\footnote{\texttt{https://github.com/SWOTAlgorithms/RiverObs}} which aggregates the WSEs from a selection of pixels to issue WSEs at nodes every 200~m along river centerline and at river reach every 10~km (which are defined by the SWOT River Database or SWORD \cite{allen2018}).
The selection of pixels is based on a class assignment related to the class of the water pixels (i.e. open water, water near land, \textit{dark water}\footnote{\textit{Dark water} refers to smooth water surfaces that produce low measurement signals due to specular reflection, wherein the satellite’s emitted signal is reflected away from the sensor rather than back toward it.}, etc.), and weighted by their respective uncertainties to provide node-level WSEs with an error below 10~cm as prescribed in the SWOT requirements.
Several factors can contribute to the occurrence of \textit{dark water} in SWOT observations, including calm water surfaces at low incidence angles, signal attenuation or dropout due to rainfall, persistent or seasonal vegetation cover, and reduced signal-to-noise ratio within the swath, which can lead to undetected water surfaces \cite{fu2024surface}.
The details for SWOT WSE, width, and slope calculation are given in  SWOT River Single Pass Product Description Document \cite{swotproduct2020} and example data products\footnote{\texttt{https://podaac.jpl.nasa.gov/swot?tab=datasets}}.

Figure~\ref{fig:SWOT_nodes} shows a typical SWOT node-level river product over a meander of a reach of the Garonne River observed on day 8 (third vertical line in Figure~\ref{fig:obs_OSSE}) of the OSSE (Figure~\ref{fig:SWOT_nodes_OSSE}) for the synthetical SWOT and on 2024-02-28 (Figure~\ref{fig:SWOT_nodes_real}) for the real one, where each node is an aggregation of pixels. The pixels are color-coded either by their associated node on the left panel, or by their class on the right panel. For each member of the ensemble, the model equivalent of the SWOT node-level WSE is computed using the SWOT-specific observation operator.  It is the average of the WSE of certain associated pixels selected according to their quality flag; each pixel WSE is interpolated from the TELEMAC-2D mesh of simulated water depths.

\subsection{Data assimilation for OSSE}

\begin{figure*}[t]
\centering
\includegraphics[width=0.8\linewidth]{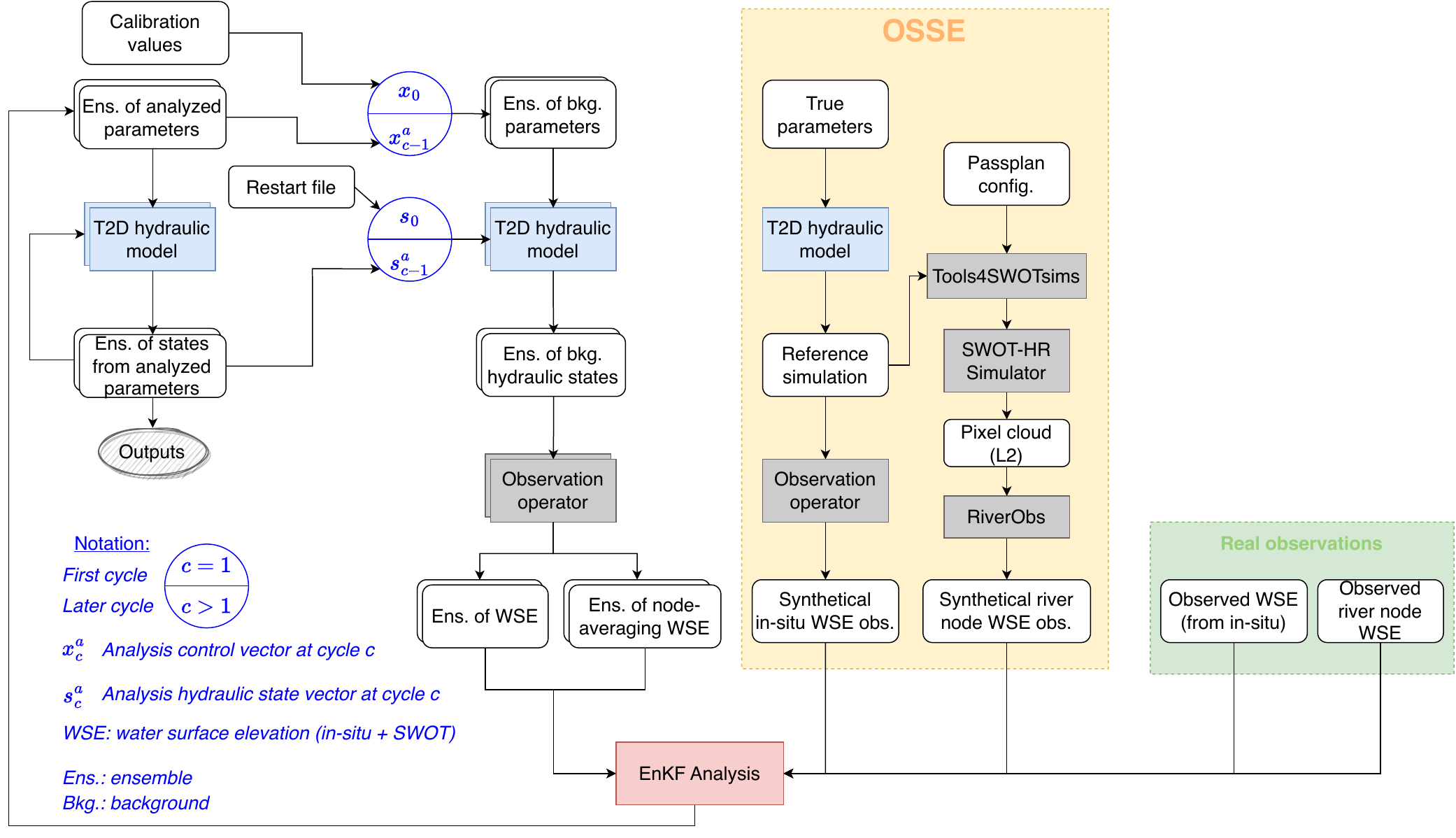}
\caption{Workflow of EnKF DA strategy in OSSE (orange box) and real event (green box).}
\label{fig:workflow}
\end{figure*}

As for flow rate $Q$, the DA corrects and updates the friction coefficients $K_s$ in order to approximate as closely as possible the measurements from in-situ and remote sensing sources. 
\huy{Figure~\ref{fig:workflow} illustrates the assimilation workflow, with key OSSE-specific processes highlighted in the orange box. The reference simulation, which serves as the basis for generating synthetic observations, is configured using parameters derived from prior assimilation of real in-situ data. Dedicated observation operators are employed to derive model equivalents of the observations for each ensemble member. True WSE values are extracted from the reference simulation at corresponding observation times and locations, and are used to generate synthetic in-situ data at the stream-gauge stations.}

\huy{Three EnKF-based experiments are carried out, as detailed in Table~\ref{tab:settings}. The deterministic simulation driven by the true control vector is referred to as the \textit{Truth}, while the simulation without assimilation, using previously calibrated parameters, is denoted as the \textit{Open Loop} (OL). The IDA experiment assimilates only in-situ WL measurements, SWDA relies solely on synthetic SWOT data, and FDA combines both observation sources. To assess assimilation performance, flood extent maps from four selected time steps---capturing the rising/falling limb and peak of the flood---are extracted from the \textit{Truth} and used as spatial reference benchmarks. The time steps are represented by vertical black-dashed lines in Figure~\ref{fig:obs_OSSE}.}


\huy{Within all three DA experiments, the control vector consists of the friction coefficients in the floodplain and riverbed ($K_{s_i}$, with $i \in [0,6]$), along with the inflow multiplicative coefficient ($\mu$). For in-situ observations, the proportionality factor $\tau$ used to define the observation error is set to 15\%, resulting in a standard deviation $\sigma_{obs}$ equal to 15\% of the corresponding observed value.}

Much shorter revisit times (i.e. 6h, 12h, 18h, and 24h) of SWOT data were also evaluated, following SWDA configuration. While these daily and sub-daily revisit times do not reflect the actual passage of the SWOT satellite, they are considered for two reasons. First, it is hypothesized that increasing the measurement frequency could help reduce modeling errors. Second, such an approach aims to assess the capacity of the assimilation strategy to integrate new space missions that offer higher temporal resolution in the future, e.g. Sentinel-3 Next Generation Topography (S3NG-TOPO) mission \cite{2023osts.confE.137E}.

\begin{table}[t]
\centering
\caption{Summary of the OL and DA experiment settings.}
\label{tab:settings}

\begin{tabular}{p{0.1\linewidth}ccll}
    \hline
    & \multicolumn{2}{c}{Assimilated observations} &  \\ \cline{2-3}
    Exp. & In-situ & SWOT & Control vector & Control vector \\ 
    name & WSE & WSE & (OSSE) & (real event) \\ \hline
    \textcolor{orange}{OL} & $\square$ & $\square$ & - & - \\
    \textcolor{green}{IDA} & $\done$ & $\square$ & $K_{s_{[0:6]}}, \mu$ & $K_{s_{[1:5]}}, \mu$ \\ 
    \textcolor{red}{SWDA} & $\square$ & $\done$ & $K_{s_{[0:6]}}, \mu$ & $K_{s_{[1:5]}}, \mu$ \\
    \textcolor{violet}{FDA} & $\done$ & $\done$ & $K_{s_{[0:6]}}, \mu$ & $K_{s_{[1:5]}}, \mu$ \\
\hline
\end{tabular}

\end{table}

\subsection{Data assimilation for real 2024 event}

As explained previously in subsection \ref{subsect:TELEMAC-2DGaronne}, the number of friction zones in the riverbed is reduced to five, whereas the total number of gauge stations is now seven with many Vortex-io micro-stations being available for the 2024 event (Table~\ref{tab:in-situ}). The study period is 45 days between 2024-02-10 and 2024-03-26. 
Figure~\ref{fig:obs_2024} depicts the observed in-situ WL for the event at VigiCrue observing stations, namely Tonneins (TON), Marmande (MD0) and La Réole (LR0), in blue, orange and green lines, respectively. It should be noted that other in-situ data at Vortex-io observing stations are also available that are not shown here.
The overpass times SWOT and Sentinel-6 are indicated as vertical blue dash-dotted lines and cyan dotted lines, respectively.
Two hydrograph peaks have been observed with more than 2,000~\Qunit~on 2024-02-27 and 2024-03-11, but they did not cause overflows into the floodplain. The control vector is now composed of five Strickler coefficients in the main channel (i.e. $K_{s_i}$ with $i \in [1,5]$) and a corrective parameter $\mu$ for the discharge. Since the river did not cause overflows, the Strickler coefficient $K_{s_0}$ in the floodplain 
is not involved in the DA experiments for the real event. 


\section{Results and Discussions}
\label{sec:result}
This Section shows the results of the DA experiments in the control space and in the observation space, in the OSSE (subsection \ref{sect:results_OSSE}) and the real event (subsection \ref{sect:results_real}). Throughout the Section, the observations (or the \textit{Truth} in the case of OSSE) are plotted in black-dashed lines, OL is plotted in orange, IDA is in green, SWDA in red, and FDA in violet for the sake of consistency. 

\subsection{Assessment metrics}\label{sect:metrics}



The accuracy of the simulated water levels ($H^m$) is evaluated against in-situ observations ($H^o$) using the root-mean-square error ($\mathrm{RMSE}$), calculated over the assimilation windows across the entire flood event. The $\mathrm{RMSE}$ is computed from the time series sampled at observation times, according to:
\begin{equation}\label{eq:RMSE}
     \mathrm{RMSE} = \sqrt{\dfrac{1}{n_{obs}} \sum_{i=1}^{n_{obs}}\left(H_i^m-H_i^o\right)^2}   
\end{equation}

The $\mathrm{RMSE}$ is also computed against Sentinel-6 WSE profiles along the river centerline for each experiment. For the 2D assessments, the Critical Success Index ($\mathrm{CSI}$) is employed as the 2D performance metric, calculated over the model domain relative to the \textit{Truth} flood extent maps. The $\mathrm{CSI}$ ranges from 0\%, indicating no spatial agreement, to 100\%, representing a perfect match between simulated and \textit{Truth} flood extents.

\subsection{Results for OSSE experiments}\label{sect:results_OSSE}

\huy{Figure~\ref{fig:control_0SSE} illustrates the evolution of the control parameters estimated across the suite of DA experiments conducted in this study, starting with Figure~\ref{sfig:control_0SSE} for the original pass plan. Specifically, the friction coefficients $K_{s_i}$ (with $i \in [0,6]$) for the riverbed and floodplain, as well as the scalar inflow correction  $\mu$, are presented. The true parameter values, from the reference simulation, are plotted in black and the calibrated coefficients are indicated by horizontal orange-dashed lines. The bottom panels of Figure~\ref{fig:control_0SSE} present the temporal evolution of the upstream BC discharge at the TON station, as modified by the analysis inflow correction factor $\mu$. This highlights the degree to which the DA framework adjusts the discharge time series to reconcile model dynamics with observations. The timing of the synthetic SWOT satellite overpasses during the flood event is indicated by vertical dashed lines. These mark the assimilation time steps at which observational updates were introduced, particularly for SWDA, enabling assessment of their impact on parameter estimation and flow dynamics.}

\huy{Table~\ref{tab:RMSE_OSSE} summarizes the 1D quantitative assessment across all OSSE experiments. The $\mathrm{RMSE}$s were computed between simulated and reference WLs at the three VigiCrue observation stations, TON, MD0 and LR0, over the whole duration of the synthetical flood event.
In addition, Table~\ref{tab:RMSE_OSSE} also reports the CSI values computed between the simulated and reference flood extent maps at four selected time steps during the event. The first two time steps (days 17 and 19) correspond to the rising limb of the flood, the third (day 20) captures the flood peak, and the final time step (day 23) is during the falling limb.}

\subsubsection{For original SWOT pass plan}
\begin{figure*}[!t]
\begin{subfigure}[b]{0.49\linewidth}
    \centering
    \includegraphics[width=\linewidth]{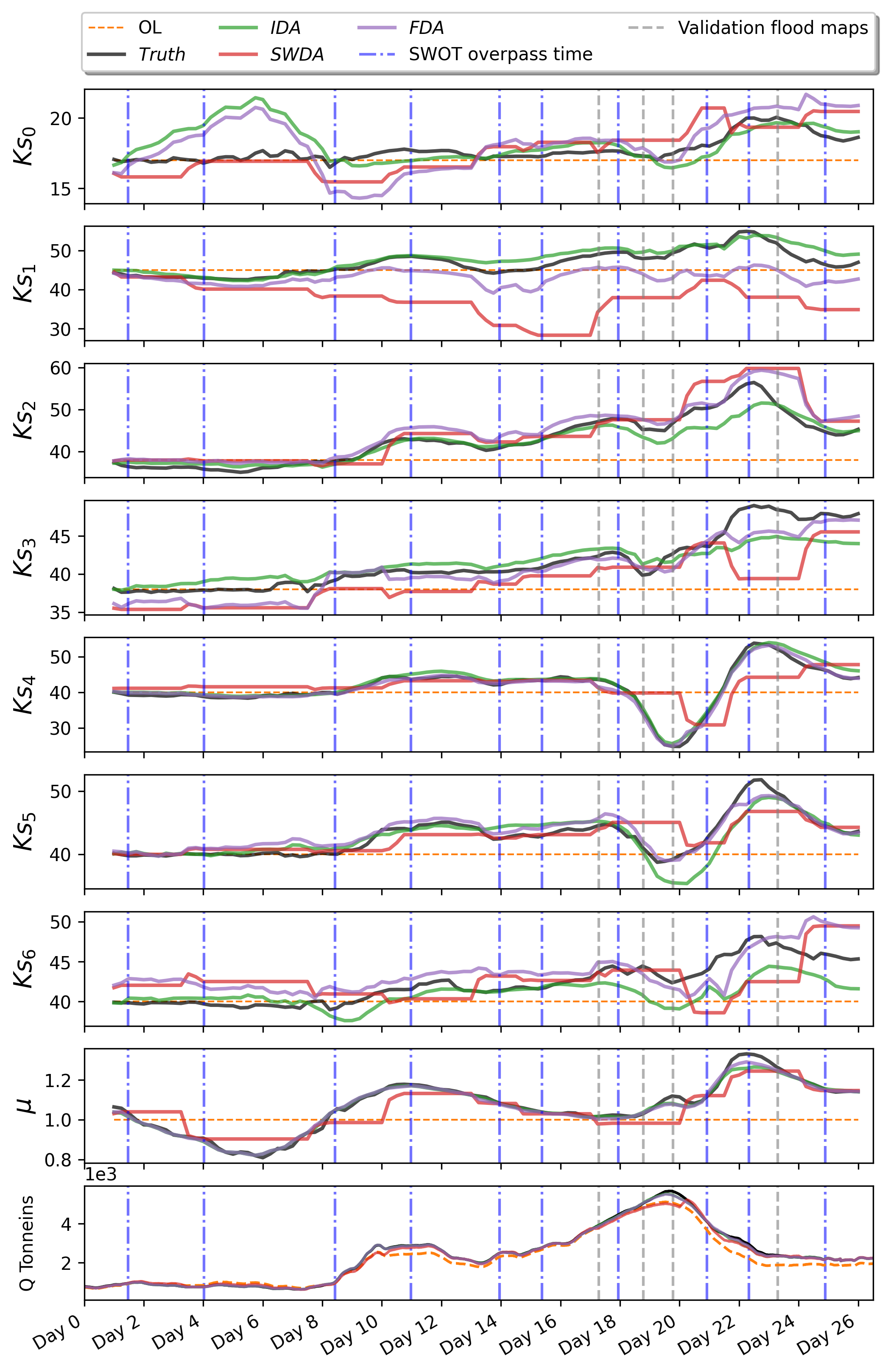}
    \caption{OSSE with original pass plan}
    \label{sfig:control_0SSE}
\end{subfigure}
\hfill
\begin{subfigure}[b]{0.49\linewidth}
    \centering
    \includegraphics[trim=0 0 0 0, clip,width=\linewidth]{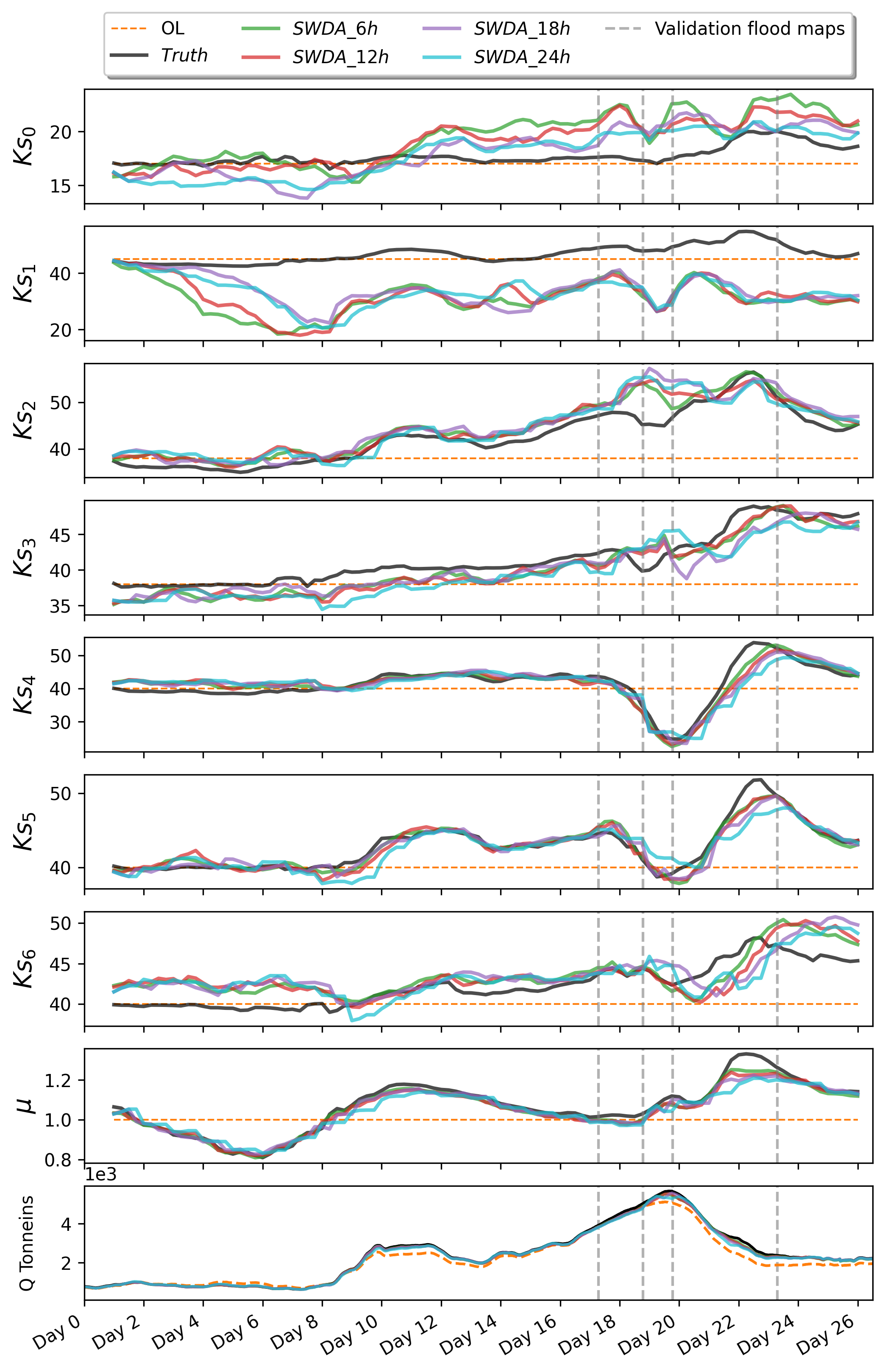}
    \caption{OSSE with shorter pass plans}
    \label{sfig:SWDA_param_compare}
\end{subfigure}
    \caption{Analyzed values of the control vector by IDA (green), SWDA (red), and FDA (violet) in (a) OSSE with the original pass plan, and by SWDA at different SWOT revisit times (b) OSSE with the shorter pass plans. 
    From top to bottom: floodplain friction coefficient $K_{s_0}$, riverbed friction coefficients $K_{s_i}$ for $i \in [1,6]$, inflow correction factor $\mu$, and upstream forcing $Q(t)$ and the simulated discharge time-series at Tonneins.}
    \label{fig:control_0SSE}
\end{figure*}


\huy{The assessment, in the control space, of the assimilation skill among the three DA experiments revealed notable differences in performance. 
The IDA experiment, which assimilates only in-situ WL observations, demonstrated high performance, recovering both the true parameter values (Figure~\ref{sfig:control_0SSE}) and the associated hydraulic states (Table~\ref{tab:RMSE_OSSE}).
The SWDA, relying exclusively on SWOT observations, was unable to retrieve the true parameter values. This emphasizes the limitation of SWDA to fully constrain the system, mainly due to the temporal resolution of the SWOT data. In contrast, the FDA experiment, which assimilated both SWOT and in-situ data, demonstrated a robustness to recover the true system states and parameters, benefiting from complementary information without substantial evidence of observational conflict. 
Overall, FDA proved to be the most effective approach among the three experiments.}


\huy{In the context of the OSSE, both IDA and FDA demonstrated the ability to retrieve the true riverbed friction coefficients, albeit with varying degrees of accuracy. This was particularly evident in the downstream segments, especially the fourth and fifth river segments, corresponding to $K_{s_4}$ and $K_{s_5}$. However, here it can be remarked that the FDA analyses (violet lines) are more closely aligned with the known reference values (black lines) than the IDA analyses (green lines). The added value of SWOT data in FDA can also be remarked over the analysis of $\mu$ during the falling limb, at day 23.
Given that the \textit{Truth} was derived from prior model reanalysis (subsection~\ref{ssec:osse}), the agreement of both IDA and FDA analyses with the \textit{Truth} can be anticipated. Nevertheless, the SWDA results (red lines) reveal the presence of equifinality in parameter estimation: comparable WL fits at observation stations can be achieved through compensating errors, such as a combination of lower $K_{s_3}$ with higher $K_{s_4}$ values, or vice versa. This underscores the inherent challenges in achieving unique parameter identification using sparse observational data.}



\huy{The presence of equifinality is also evident in the early-stage (before day 8) analysis of floodplain friction $K_{s_0}$ in IDA and FDA. As water begins to spread across the floodplain, this equifinality diminishes, allowing the floodplain friction estimates to progressively align with the true value, particularly by IDA. Despite these compensatory effects, all analyzed friction coefficients, both within the channel and across the floodplain, remain within physical ranges and are consistently closer to the true parameter values than to the default value, especially near peak flow.}

\huy{Moreover, the assimilation of in-situ WL observations at Marmande (MD0), located in the fourth river segment, enables highly precise retrieval of $K_{s_4}$ in the IDA and FDA. Here, additional information from SWOT data, in FDA, only slightly improve the parameter estimation during the flood recess (after day 22), indicating that local in-situ data alone can effectively constrain the friction. On the other hand, the FDA analyses for $K_{s_3}$ and $K_{s_5}$ outperform those from IDA, highlighting the advantage of SWOT observations over these ungauged river segments.}

\begin{figure}[t]
\centering
\begin{subfigure}[b]{0.97\linewidth}
\centering
\includegraphics[trim=0 0 0 0, clip,width=0.99\linewidth]{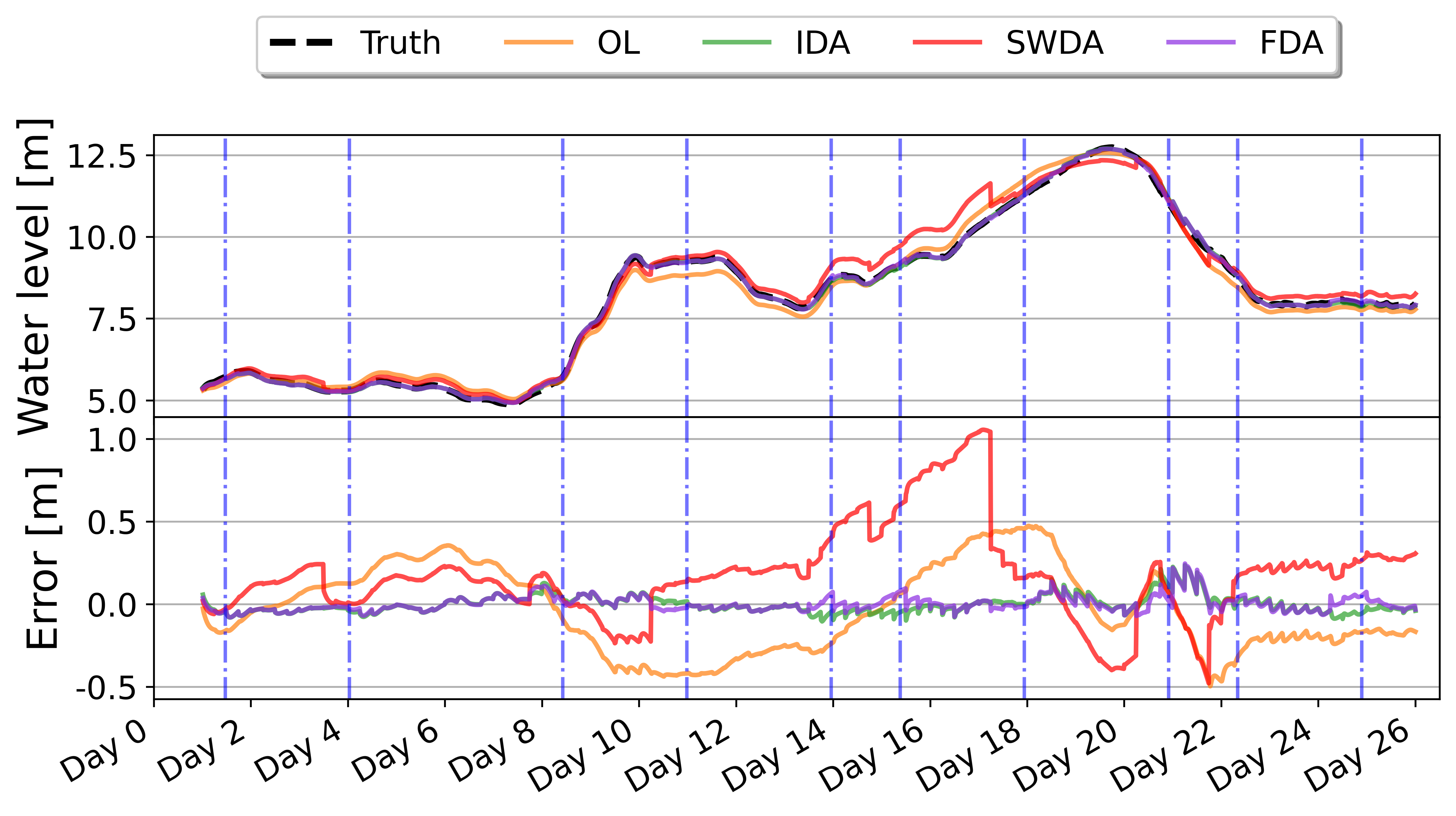}
\caption{Tonneins (TON)}\label{fig:H_Tonneins_OSSE}
\end{subfigure}

\begin{subfigure}[b]{0.97\linewidth}
\centering
\includegraphics[trim=0 0.3cm 0 1.4cm, clip,width=0.99\linewidth]{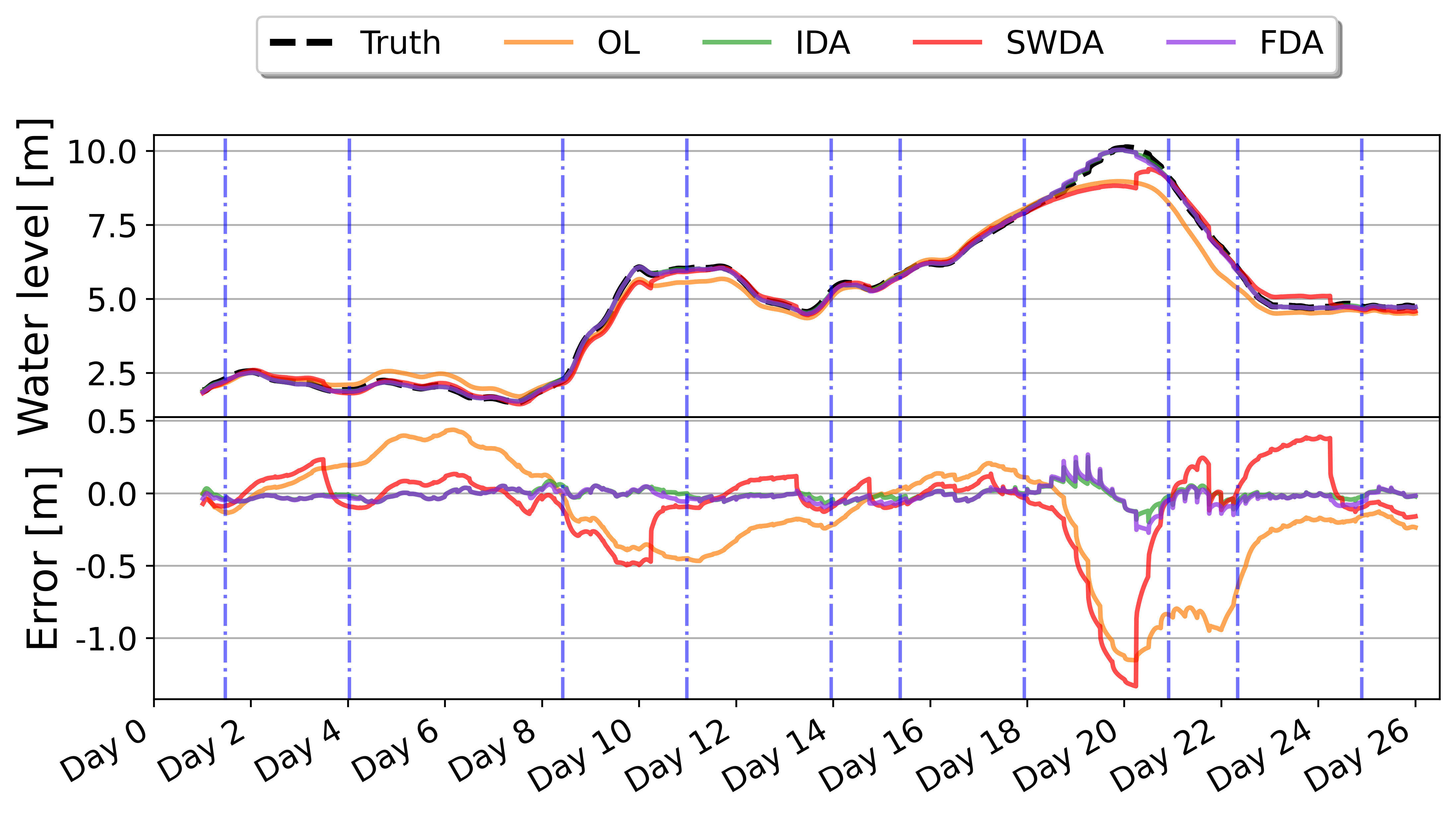}
\caption{Marmande (MD0)}\label{fig:H_Marmande_OSSE}
\end{subfigure}

\begin{subfigure}[b]{0.97\linewidth}
\centering
\includegraphics[trim=0 0.3cm 0 1.4cm, clip,width=0.99\linewidth]{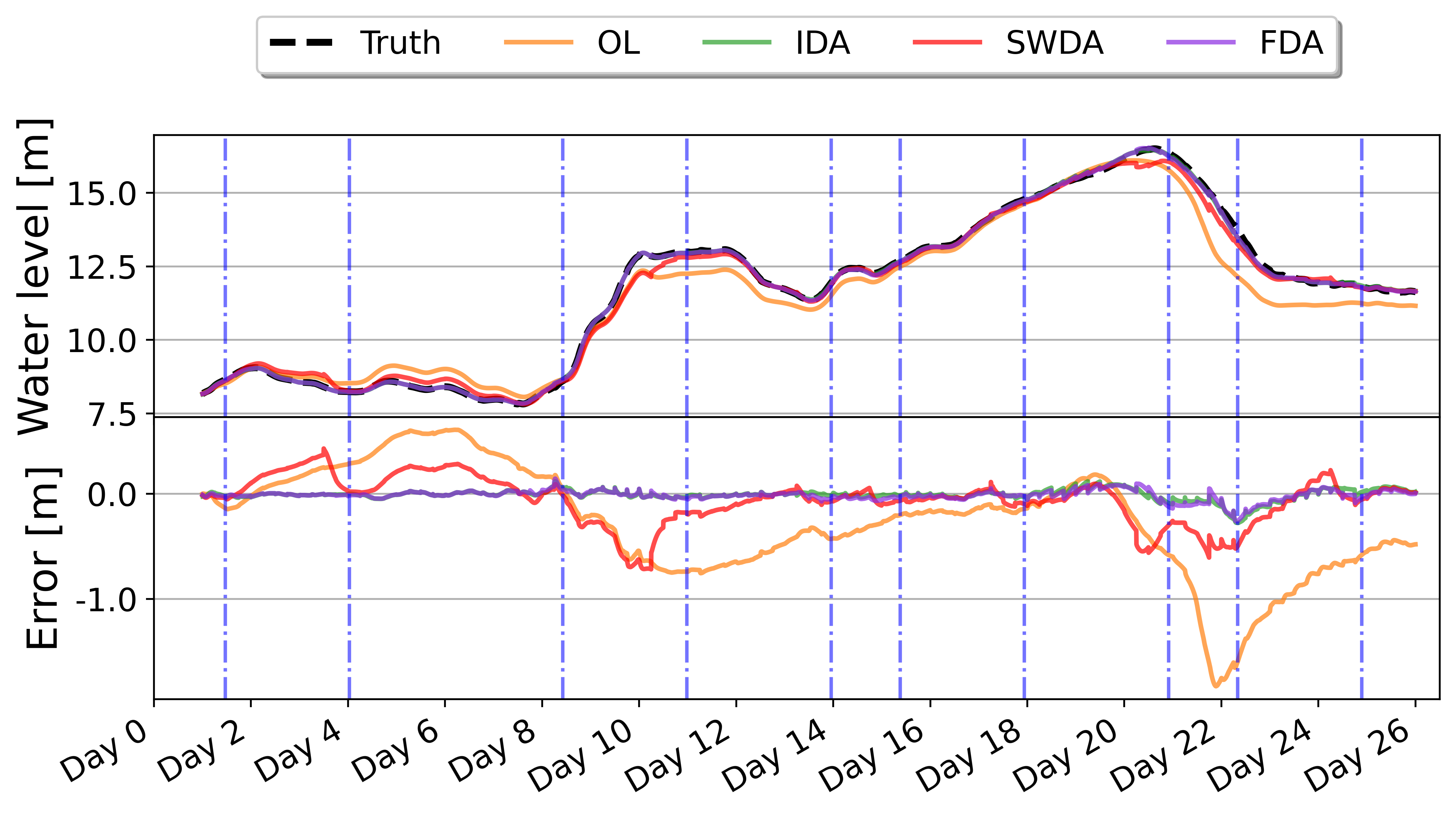}
\caption{La Réole (LR0)}\label{fig:H_LaReole_OSSE}
\end{subfigure}
\caption{WLs at observing stations (a) Tonneins, (b)  Marmande and (c) La Réole for all experiments (top panels), and errors computed with respect to the \textit{Truth} in-situ WLs (bottom panels).}
\label{fig:H_OSSE}
\end{figure}

\huy{Figure~\ref{fig:H_OSSE} presents the simulated WLs at the TON (Figure~\ref{fig:H_Tonneins_OSSE}), MD0 (Figure~\ref{fig:H_Marmande_OSSE}), and LR0 (Figure~\ref{fig:H_LaReole_OSSE}) stations for the OL and all three DA experiments, alongside the \textit{Truth} WLs indicated by black dashed lines. The OL simulation exhibits a slight under-estimation of the flood peak across the domain. In contrast, the DA strategies demonstrate a marked improvement in correcting flow propagation within the river channel.}

\huy{At the TON station (Figure~\ref{fig:H_Tonneins_OSSE}), a discrepancy of 1~m was observed during the rising limb, highlighting once again the sensitivity of model accuracy to observation timing. This error was substantially reduced when the DA system assimilated the SWOT observation on day 18 (seventh vertical line in Figure~\ref{fig:H_OSSE}), reducing the WL errors to approximately 20~cm. 
The influence of observational timing is further illustrated at the MD0 station (Figure~\ref{fig:H_Marmande_OSSE}), where the flood peak is underestimated by SWDA, as it occurred between two SWOT overpasses on day 18 and 21. Without timely observational updates, the model—constrained by previously estimated parameters—fails to capture the sharp rise in discharge.}

\huy{Regarding the 2D assessment of simulated flood extents compared to the \textit{Truth} maps, Table~\ref{tab:RMSE_OSSE} demonstrates that all DA experiments, especially IDA and FDA, exhibit improved performance and achieve higher $\mathrm{CSI}$ values relative to the OL simulation. Specifically, while the FDA provides superior results during the rising limb, the IDA attains the highest $\mathrm{CSI}$ at the flood peak on day 20. Although SWDA does not match the same level of WL accuracy at individual gauge stations, it nevertheless provides a strong representation of spatial flood dynamics, particularly during the rising and falling limbs. However, the absence of SWOT observations precisely at the flood peak limits the capability of SWDA during this critical phase, resulting in lower $\mathrm{CSI}$ values on day 20.}

\begin{table*}[!ht]
\centering
\caption{$\mathrm{RMSE}$ [m] of simulated WLs computed with respect to \textit{Truth} in-situ WLs at observing stations and $\mathrm{CSI}$ [\%] of simulated flood extent maps with respect to \textit{Truth} flood extent maps. The lowest $\mathrm{RMSE}$ and highest $\mathrm{CSI}$ among the experiments for each column is boldfaced.}
\label{tab:RMSE_OSSE}
\begin{tabular}{c|ccc|cccc}
\hline
& \multicolumn{3}{|c|}{$\mathrm{RMSE}$ against in-situ \textit{Truth} WLs} & \multicolumn{4}{c}{$\mathrm{CSI}$ against \textit{Truth} flood maps} \\ \hline
Exp. & TON & MD0 & LR0 & Day 17 (rising) & Day 19 (rising) & Day 20 (peak) & Day 23 (recess) \\ \hline
\textcolor{orange}{OL} & 0.260 & 0.398 & 0.578 & 87.95\% & 81.58\% & 80.64\% & 77.52\% \\
\textcolor{green}{IDA} & 0.052 & \textbf{0.042} & 0.053  & 92.89\% & 81.79\% & \textbf{90.77\%} & 76.57\% \\ 
\textcolor{red}{SWDA} & 0.325 & 0.291 & 0.229 & 91.69\% & 82.15\% & 73.52\% & \textbf{80.27\%} \\
\textcolor{violet}{FDA} & \textbf{0.047}& 0.056& \textbf{0.051} & \textbf{93.07\%} & \textbf{82.44\%} & 88.32\% & 78.92\% \\ 
\hline
SWDA 6h & 0.450 & 0.083 & \textbf{0.104} & 92.39\% & 81.12\% & \textbf{95.25\%} & 76.37\% \\
SWDA 12h & 0.428 & \textbf{0.082} & 0.115 & 92.21\% & 81.10\% & 93.47\% & 76.52\% \\
SWDA 18h & \textbf{0.350} & 0.087 & 0.132 & 92.43\% & \textbf{81.92\%} & 91.46\% & 76.52\% \\
SWDA 24h & 0.368 & 0.110 & 0.191 & \textbf{92.50\%} & 81.51\% & 90.30\% & \textbf{78.34\%} \\
\hline
\end{tabular}
\end{table*}


\subsubsection{For shorter pass plans in OSSE}\label{sssec:osse_new pass plan}

\huy{As previously discussed, the performance of the SWDA  is strongly limited by the revisit frequency of SWOT, with current intervals proving insufficient to match the effectiveness of in-situ observations. Leveraging the OSSE framework, additional SWDA experiments were conducted to evaluate the potential benefits of future altimetry missions offering similar observational capabilities but with increased temporal resolution. Specifically, DA experiments were repeated under hypothetical revisit scenarios of 6, 12, 18, and 24 hours, as described earlier.}



\huy{It was shown that the SWDA experiments with increasingly denser SWOT observations, are able to retrieve the \textit{Truth} parameter values. 
Accordingly, as the revisit frequency decreases from 24 hours to 6 hours, the SWDA experiments produce analyzed parameter values that progressively converge toward the \textit{Truth}, approaching the accuracy achieved by IDA under the original pass plan (green lines in Figure~\ref{sfig:control_0SSE}).
This is mostly true for all parameters, except $K_{s_0}$ and $K_{s_1}$. 
While this could be expected for $K_{s_0}$, the analyzed values of $K_{s_1}$ by different SWDA experiments reveal some problem with SWOT synthetical data. 
Indeed, regarding the floodplain friction $K_{s_0}$, the IDA and FDA in the previous experiments using the original pass plan also manifested the same pattern. In contrast, for $K_{s_1}$, the assimilation of SWOT observations---particularly those near the first meander of the TELEMAC-2D model domain after Tonneins, where the riverbed is governed by $K_{s_1}$---reveals signs of overbank flow into the adjacent floodplain. To maintain water levels consistent with these SWOT observations, both $K_{s_0}$ and $K_{s_1}$ are reduced, effectively delaying water evacuation from the floodplain and preserving inundation in this area over an extended period.}

\huy{These results underscore the advantages of increased revisit frequency in future satellite missions. The setups with a  6-hour and 12-hour revisit interval enable accurate retrievals of \textit{Truth} parameter values and hydraulic states (except at the TON station), almost comparable to that achieved using high temporal sampling in-situ data (15-minute intervals, as summarized in Table~\ref{tab:in-situ}) by IDA. Notably, across all four shorter pass plans, the impact on the discharge simulation is substantial, as evidenced by the close agreement between the reference discharge (black line) and the reconstructed time-series following the application of the corrected $\mu$ values in the SWDA experiments (bottom panel in Figure~\ref{sfig:SWDA_param_compare}).}

\huy{The key outcome of the OSSE results lies in the validation of assimilation robustness under known conditions, enabling a direct comparison against the \textit{Truth} parameter values. The limitations of the SWDA experiment, due to the coarse temporal resolution of SWOT observations, were highlighted. This is reflected in the degraded estimation of friction parameters, notably during the flood peak (Figure~\ref{sfig:control_0SSE}). This has been significantly reduced when the pass plan is enhanced with shorter revisit frequency. Comparing the SWDA under the original pass plan (one SWOT every 2-3 days) to the SWDA with a 18-hour revisit frequency, the $\mathrm{RMSE}$ at the MD0 and LR0 observing stations decreases by 70.1\% and 42.4\%, respectively. It can also be asserted that revisit intervals shorter than 18 hours begin to exhibit diminishing returns, offering no clear advantage relative to the increased volume of assimilated observations (in sub-daily pass plans).}

\huy{It is notable that the $\mathrm{CSI}$ values obtained from all four SWDA experiments with shorter pass plans are comparable (on day 17, 19 and 23) to or exceed (on day 20) those achieved by IDA under the original pass plan. This highlights the substantial impact of SWOT revisit frequency on $\mathrm{CSI}$ performance, particularly evident during the flood peak on day 20, where the SWDA with a 6-hour revisit interval attains the highest $\mathrm{CSI}$ (i.e. 95.25\%). These findings underscore the value of SWOT observations in enhancing the 2D representation of floodplain dynamics, while maintaining the accuracy of simulated WLs within the river channel (except at TON).}


\subsection{Results for real experiment}\label{sect:results_real}

\huy{As previously outlined, the DA methods evaluated through OSSEs were subsequently applied to real SWOT observations acquired during the 2024 event. However, the prevailing meteorological conditions bring a major difference that is no overflowing in floodplain was observed in 2024. While the real SWOT data are not to be validated under flooded conditions, it is still relevant to assess their accuracy within the main channel. 
Unlike in OSSE, the \textit{Truth} parameter values are not known beforehand in the real event, precluding direct comparison against a reference.  Therefore, the validations of the DA experiments rely on the use of in-situ, SWOT and Sentinel-6 altimetry data.}

\subsubsection{Impacts on control vector}

\begin{figure*}[!t]
\centering
    \includegraphics[width=0.5\linewidth]{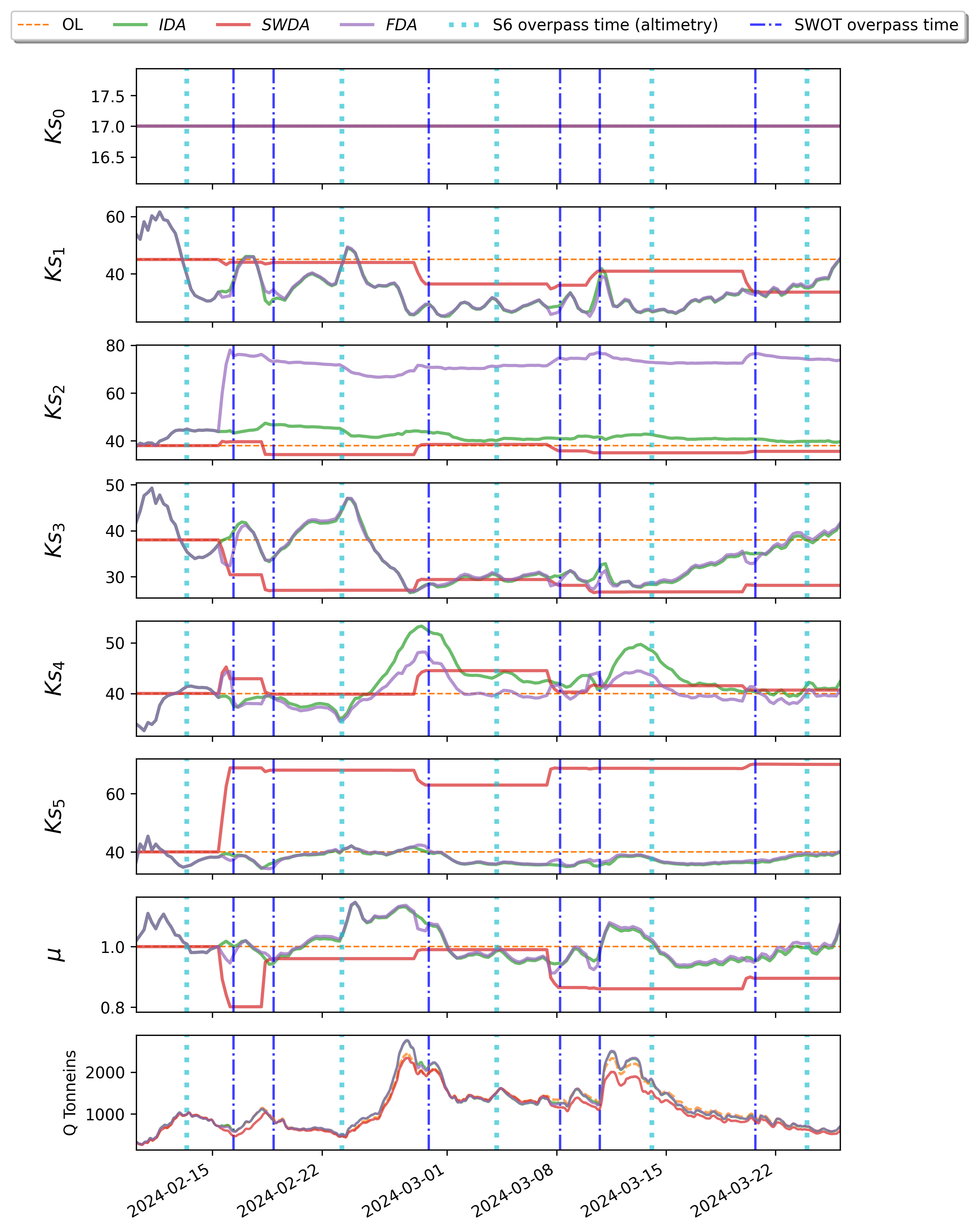}
     \caption{Analyzed values of the control vector IDA (green), SWDA (red), and FDA (violet) in real 2024 event. Horizontal orange-dashed lines represent the default parameter values, while vertical dashed lines indicate SWOT and Sentinel-6 overpass times. From top to bottom: floodplain friction coefficient $K_{s_0}$ (not controlled), riverbed friction coefficients $K_{s_i}$ for $i \in [1,5]$, inflow correction factor $\mu$, and upstream forcing $Q(t)$ and the simulated discharge time-series at Tonneins.}
     \label{fig:control_2024}
\end{figure*}

\huy{As in Figure~\ref{fig:control_0SSE}, the evolution of analyzed parameters from the different DA experiments for the 2024 real event is shown in Figure~\ref{fig:control_2024}. Horizontal orange dashed lines indicate the default or calibrated parameter values used in the OL, while green, red, and violet lines represent the IDA, SWDA, and FDA, respectively. Figure~\ref{fig:control_2024} shows the temporal variation of the friction coefficients $K_{s_i}$ for $i \in [1,5]$ (resulted from the zoning refinement of the riverbed). It should be noted that floodplain friction $K_{s_0}$ is not included as a control variable in this real-data scenario. The bottom panel in Figure~\ref{fig:control_2024} displays the upstream BC and the corresponding simulated discharge time series at the TON station, after applying the inflow correction factor $\mu$.}


\huy{The parameter estimates obtained from the IDA experiment remain within physically plausible bounds.
In the case of SWDA, the analyzed parameters tend to remain closer to their initial values for a longer period due to the lower frequency of assimilation updates. 
Some notable deviations emerge when SWOT data is assimilated. For instance, the analyzed $K_{s_2}$ from FDA, or $K_{s_5}$ from SWDA, both can reach 75~\Ksunit.}
\huy{It should be reiterated here that the resolution of the inverse problem using the EnKF remains subject to equifinality, whereby different combinations of control parameters (here, $K_{s_2}$ and $K_{s_5}$) may yield similar WL outcomes. As a result, the analyzed parameter values may exhibit compensatory behavior to reproduce the expected WLs, from the observing stations.}


\huy{In this real event, with $K_{s_0}$ excluded from the control vector due to the absence of floodplain overflow, the first riverbed friction coefficient $K_{s_1}$ exhibits more stable behavior, showing reduced variability over the course of the event, particularly when compared to IDA and FDA. This contrasts with the OSSE scenario, where $K_{s_0}$ and $K_{s_1}$ frequently compensated for one another in response to equifinality effects.
It is also worth-noting that while both IDA and FDA require a higher inflow discharge (with analyzed $\mu$ from IDA and FDA are mostly more than 1), SWDA consistently constrains the inflow throughout the event, beginning from the first assimilation of SWOT observations. The contrasting behavior observed between the upstream control variables ($\mu$ and $K_{s_1}$) and the downstream parameter ($K_{s_5}$) suggests that, through the assimilation of SWOT data, the SWDA tends to reduce the inflow volume while simultaneously favoring a more rapid outflow from the hydraulic domain.}

\subsubsection{Time evolution of WLs at observing stations}
\label{sssec:wl_real}

\begin{table*}[h]
\centering
\caption{$\mathrm{RMSE}$ between (a) simulated WLs with respect to observed in-situ WLs over the 2024 event, and simulated WSEs along the river with respect to (b) SWOT and (c) Sentinel-6 WSE measurements. The lowest $\mathrm{RMSE}$ among the experiments for each column is boldfaced.}
\begin{subtable}[t]{\textwidth}
    \centering
    \caption{$\mathrm{RMSE}$ [m] of (a) simulated WLs with respect to observed in-situ WLs. Italicized columns denote gauge stations only for validation in IDA and FDA.}
    \label{tab:RMSE_real}
    \begin{tabular}{c|c|c|cc|c|cc}
    \hline
    Exp. & TON & \huy{LMA} & MD0  & \textit{MD1} & COU & \huy{\textit{LR1}} & LR0 \\ \hline
    Stricker zone & $K_{s_1}$ & $K_{s_2}$ & $K_{s_3}$ & $K_{s_3}$ & $K_{s_4}$ & $K_{s_5}$ & $K_{s_5}$ \\ \hline
    \textcolor{orange}{OL} & 0.449 & 0.098 & 0.301 & \textit{0.213} & 0.222 & \textit{0.224} & 0.195 \\
        \textcolor{green}{IDA} & \textbf{0.116} & \textbf{0.062} &  \textbf{0.065} & \textit{0.172} & \textbf{0.073} & \textit{0.152} & \textbf{0.068} \\

    \textcolor{red}{SWDA} & 0.491 & 0.201 &  0.261 & \textit{0.203} & 0.346 & \textit{0.399} & 0.383 \\
        \textcolor{violet}{FDA} & 0.127 & 0.079 &  0.073 & \textbf{\textit{0.167}} & 0.100 & \textbf{\textit{0.147}} & 0.080 \\ 

    \hline
    \end{tabular}
\end{subtable}
\vspace{0.2cm}

\begin{subtable}[t]{\textwidth}
    \centering
    \caption{$\mathrm{RMSE}$ [m] of simulated WSEs with respect to SWOT-node WSEs.}
    \label{tab:rmse_SWOT}
    \begin{tabular}{cccccccc}
    \hline
    Date & 2024-02-16 & 2024-02-18 & 2024-02-28 & 2024-03-08 & 2024-03-10 & 2024-03-20 & \\
    time & 08:26:29 & 21:45:26 & 20:07:43 & 05:11:32 & 18:30:30 & 16:52:45 & Avg. \\\hline
    Pass & 42 & 113 & 391 & 42 & 113 & 391 & - \\\hline
    \textcolor{orange}{OL} & 0.910 & 0.631 & 0.506 & 0.713 & 0.663 & 0.568 & 0.665 \\
    \textcolor{green}{IDA} & 0.855 & 0.486 & 0.536 & 0.641 & 0.589 & 0.499 & 0.601 \\
    \textcolor{red}{SWDA} & \textbf{0.562} & \textbf{0.371} & \textbf{0.260} & \textbf{0.372} & \textbf{0.238} & \textbf{0.295} & 0.350 \\
    \textcolor{violet}{FDA} & 0.578 & 0.391 & 0.292 & 0.421 & 0.254 & 0.302 & 0.373 \\
    \hline
    \end{tabular}
\end{subtable}
\vspace{0.2cm}

\begin{subtable}[t]{\textwidth}
    \centering
    \caption{$\mathrm{RMSE}$ [m] of simulated WSEs with respect to Sentinel-6 WSE measurements.}
    \label{tab:rmse_S6}
    \begin{tabular}{ccccccc}
    \hline
    Date & 2024-02-13 & 2024-02-23 & 2024-03-04 & 2024-03-14 & 2024-03-24 &  \\
    time & 08:07:14 & 06:05:45 & 04:04:16 & 02:02:48 & 00:01:21 & Avg. \\\hline
    \textcolor{orange}{OL} & 0.289 & 0.321 & 0.340 & 0.347 & 0.330 & 0.325 \\
    \textcolor{green}{IDA} & 0.318 & 0.316 & 0.234 & \textbf{0.206} & 0.329 & 0.281 \\
    \textcolor{red}{SWDA} & \textbf{0.270} & \textbf{0.307} & 0.255 & 0.420 & \textbf{0.293} & 0.309 \\
    \textcolor{violet}{FDA} & 0.318 & 0.316 & \textbf{0.233} & \textbf{0.206} & 0.329 & 0.280 \\
    \hline

    \end{tabular}
\end{subtable}
\end{table*}
 
\begin{figure*}[!t]
\centering
\begin{subfigure}[b]{0.475\linewidth}
\centering
\includegraphics[trim=0 0.3cm 0 0, clip,width=0.99\linewidth]{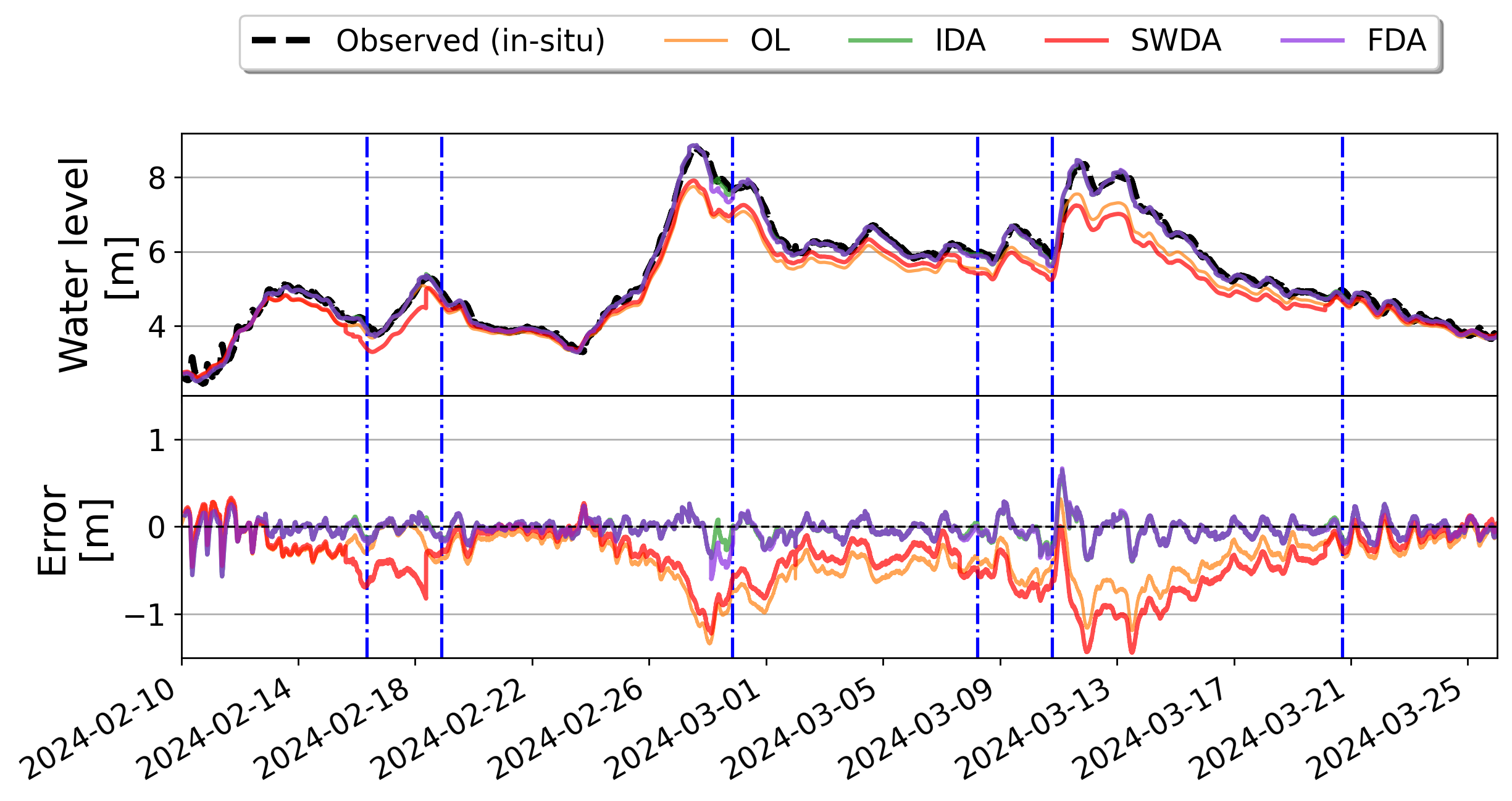}
\caption{Tonneins (TON)}
\label{fig:H_Tonneins}
\end{subfigure}
\hfill
\begin{subfigure}[b]{0.475\linewidth}
\centering
\includegraphics[trim=0 0.3cm 0 0, clip,width=0.99\linewidth]{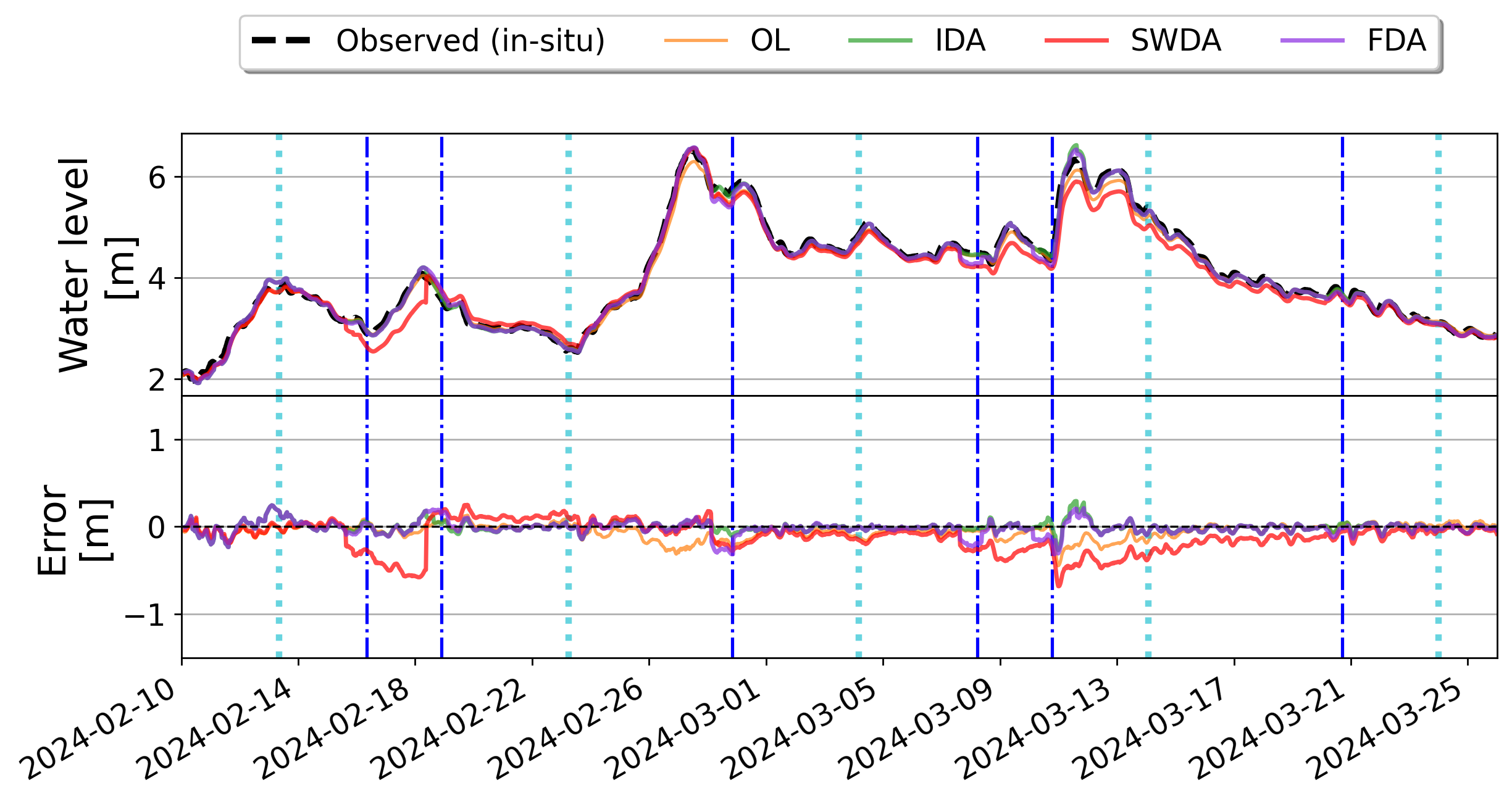}
\caption{Le Mas-d'Agenais (\huy{LMA})}
\label{fig:H_Mas_Agenais1}
\end{subfigure}

\begin{subfigure}[b]{0.475\linewidth}
\centering
\includegraphics[trim=0 0.3cm 0 1.4cm, clip,width=0.99\linewidth]{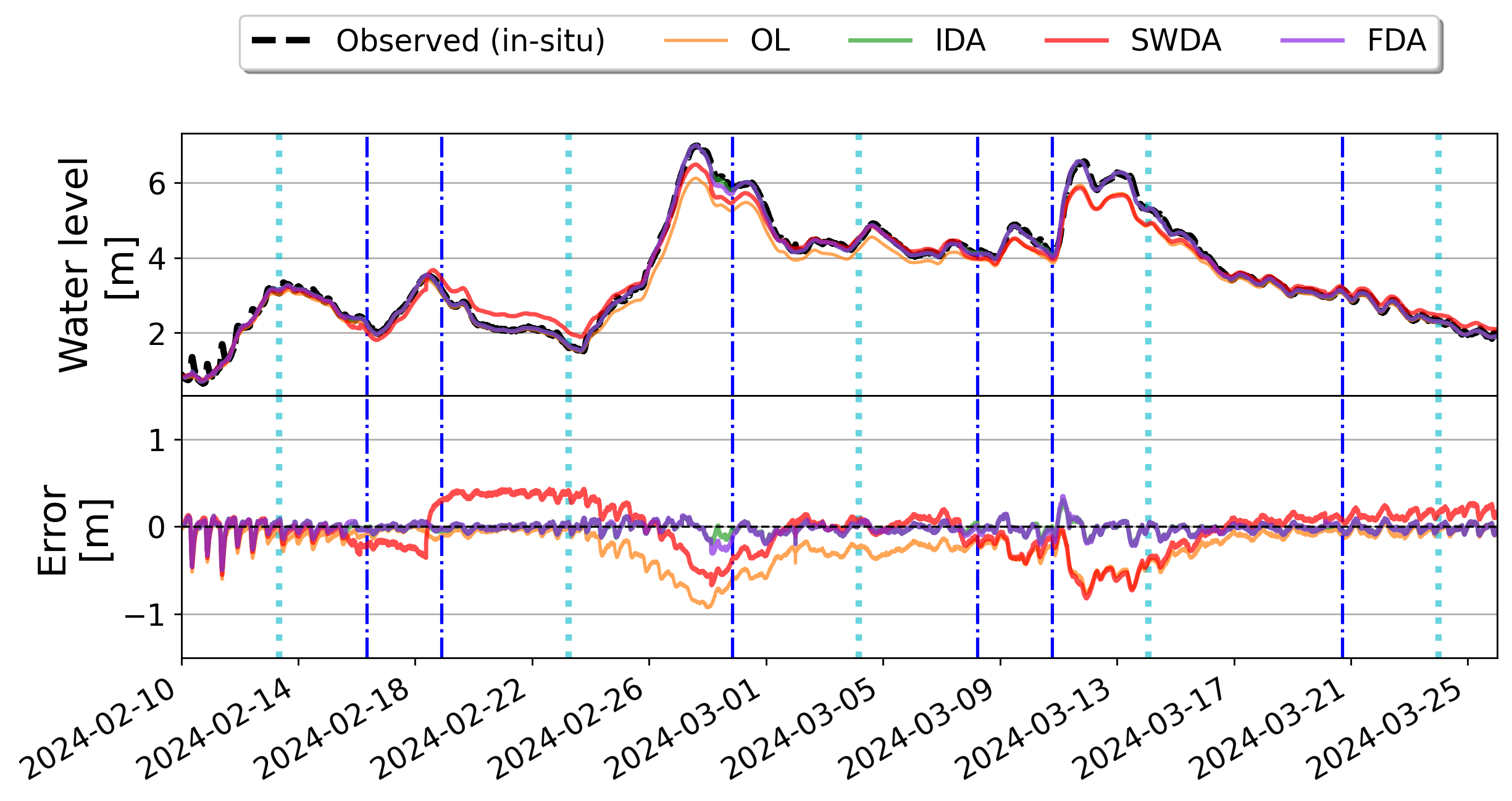}
\caption{Marmande (MD0)}
\label{fig:H_Marmande}
\end{subfigure}
%
\hfill
\begin{subfigure}[b]{0.475\linewidth}
\centering
\includegraphics[trim=0 0.3cm 0 1.4cm, clip,width=0.99\linewidth]{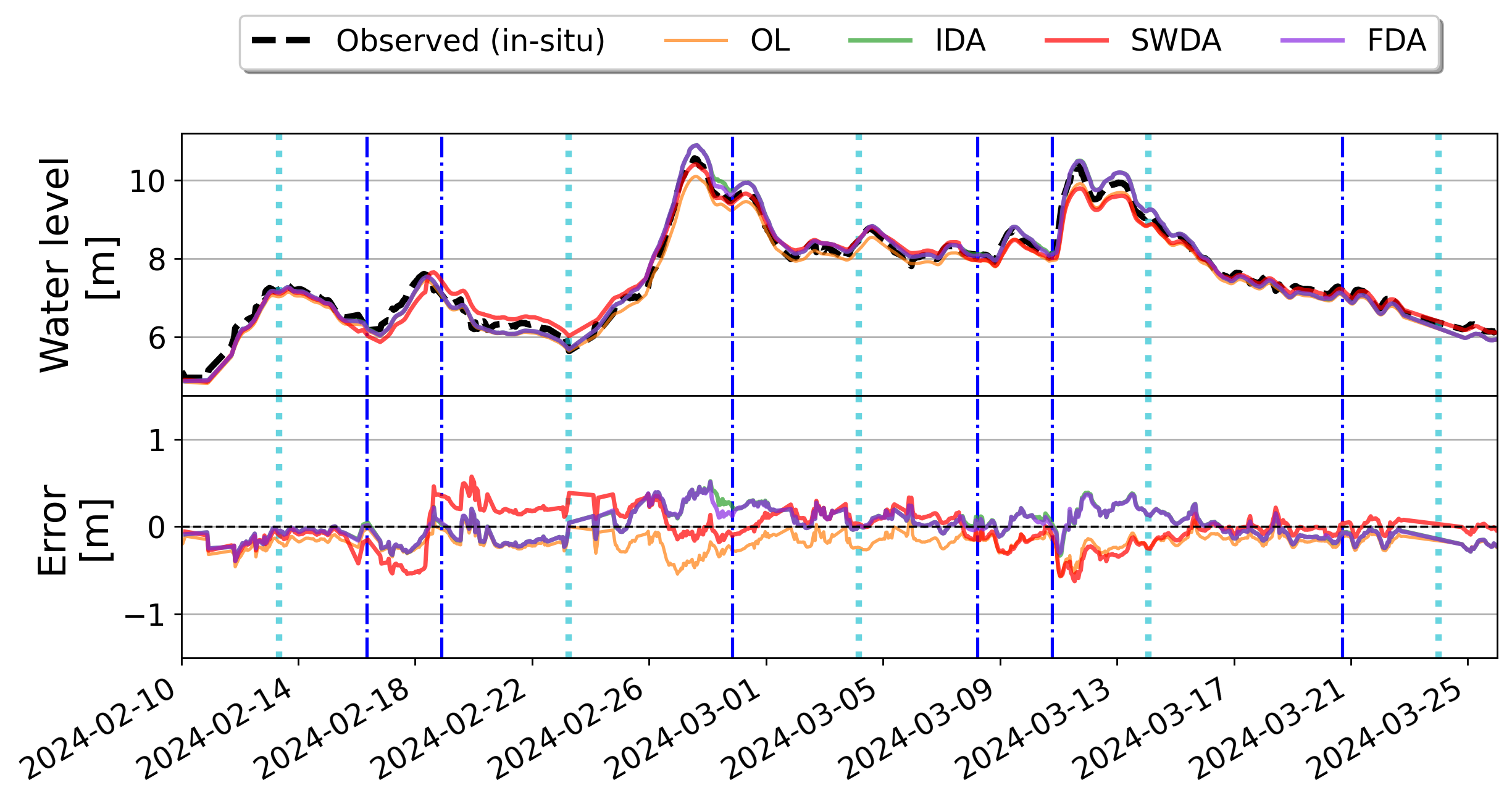}
\caption{\textit{Marmande 1 (MD1)}}
\label{fig:H_Marmande1_Vortex}
\end{subfigure}

\begin{subfigure}[b]{0.475\linewidth}
\centering
\includegraphics[trim=0 0.3cm 0 1.4cm, clip,width=0.99\linewidth]{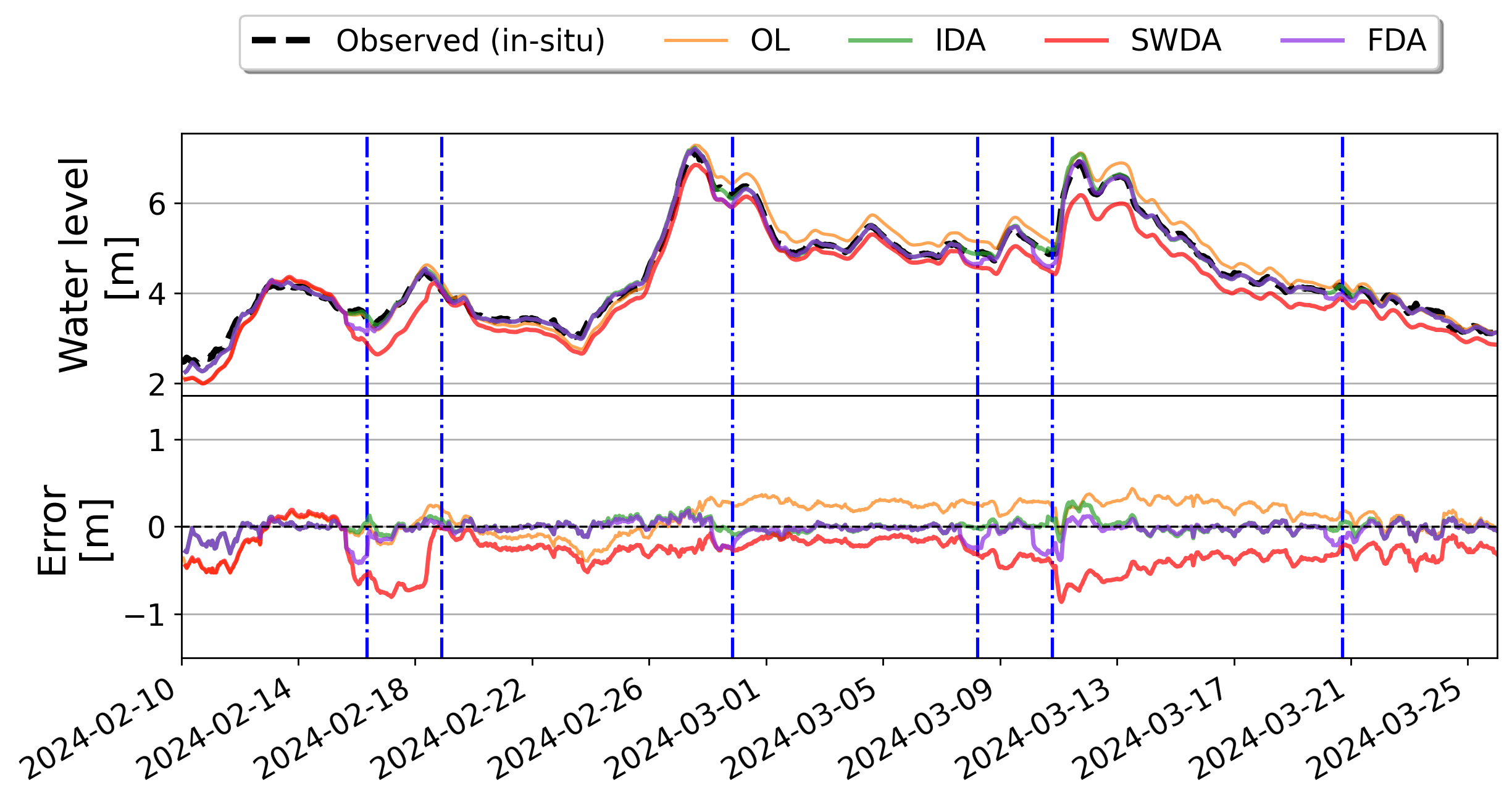}
\caption{Couthures-sur-Garonne (COU)}\label{fig:H_Couthures}
\end{subfigure}
\hfill
\begin{subfigure}[b]{0.475\linewidth}
\centering
\includegraphics[trim=0 0.3cm 0 1.4cm, clip,width=0.99\linewidth]{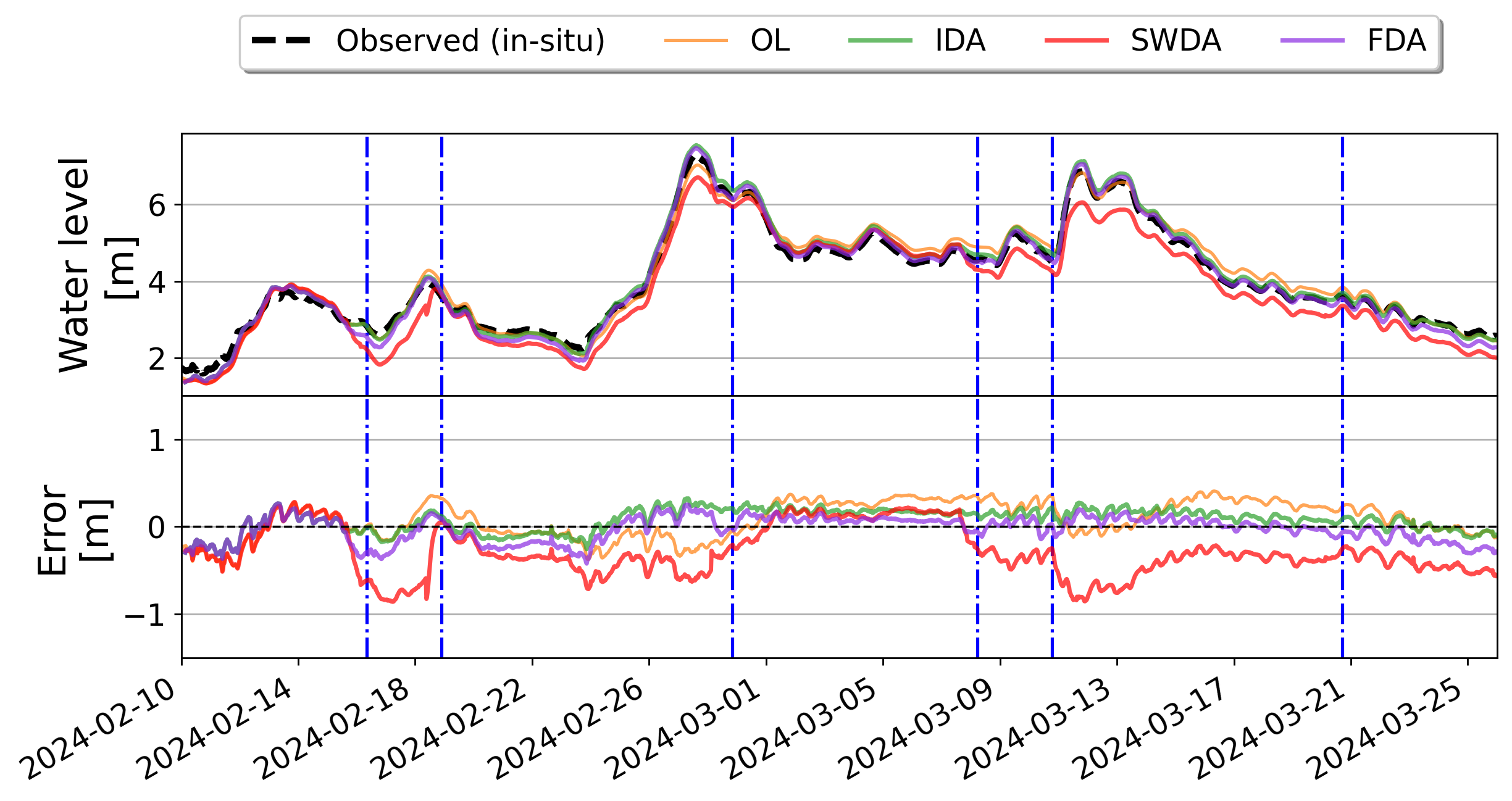}
\caption{\textit{La Réole 1 (LR1)}}\label{fig:H_LaReoleVortex}
\end{subfigure}

\begin{subfigure}[b]{0.475\linewidth}
\centering
\includegraphics[trim=0 0.3cm 0 1.4cm, clip,width=0.99\linewidth]{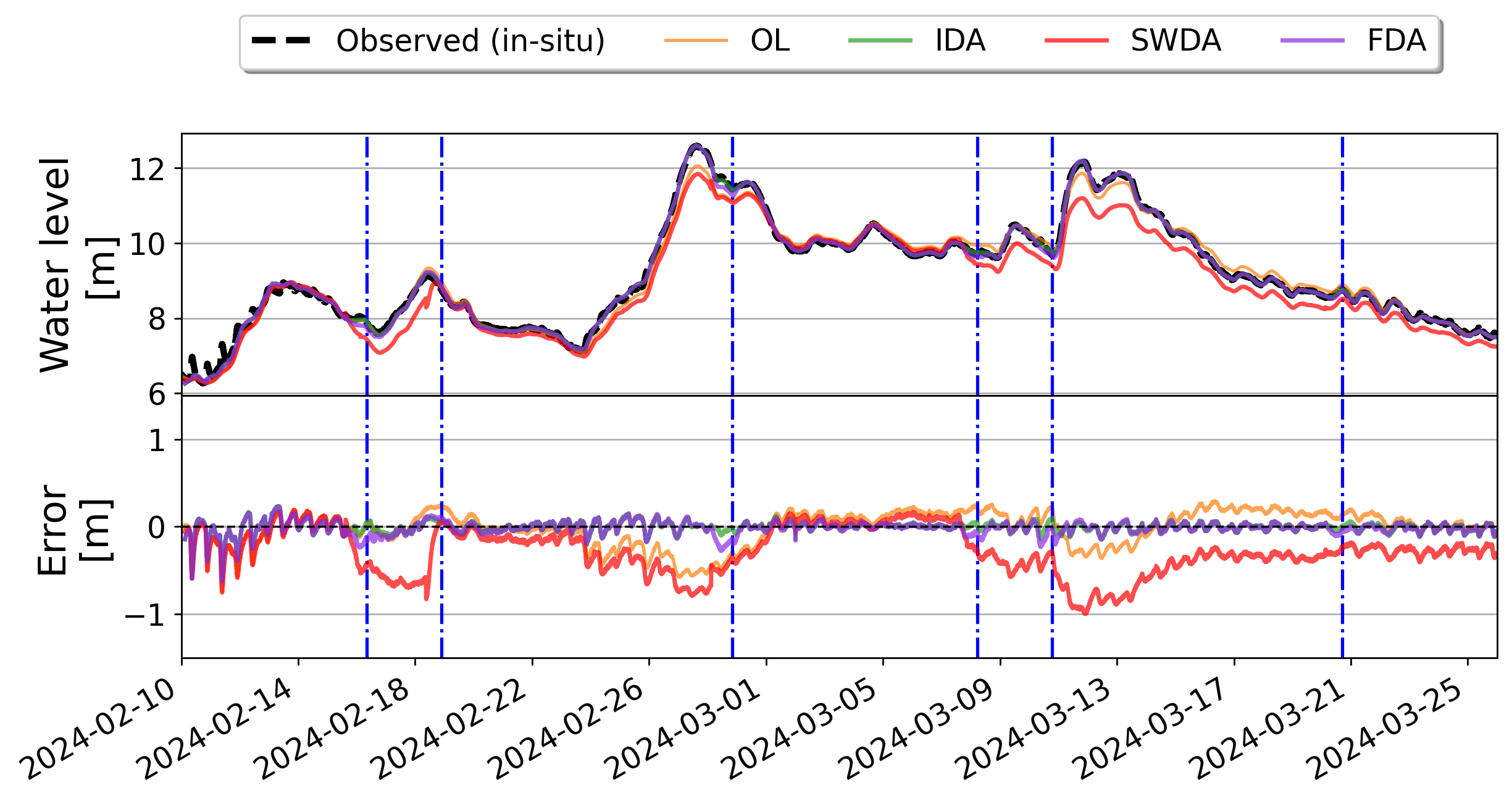}
\caption{La Réole (LR0)}\label{fig:H_LaReole}
\end{subfigure}
\hfill
\begin{subfigure}[b]{0.475\linewidth}

\end{subfigure}
\caption{WLs at observing stations for all experiments (top panels), and errors computed with respect to the observed in-situ WLs (bottom panels).}
\label{fig:H_tot}
\end{figure*}

\huy{The performance of the DA strategies was evaluated at gauge locations. 
Table~\ref{tab:RMSE_real} summarizes the $\mathrm{RMSE}$ values computed between the simulated and observed WLs at each observing station over the full duration of the event. The italicized columns indicate stations used exclusively for validation, while the remaining stations were included in the assimilation for IDA and FDA.
Figure~\ref{fig:H_tot} compares the observed (black-dashed lines) and simulated WLs (colored lines) by the OL and DA analyses.
Across all stations, both IDA and FDA experiments effectively capture the temporal evolution of WLs, with $\mathrm{RMSE}$s generally below 17~cm. At all observation stations, the $\mathrm{RMSE}$ values obtained from FDA are, in general, slightly higher (i.e. 0.8 to 2.7~cm) than those from IDA. This suggests a potential conflict or inconsistency, which can be minor yet not negligible, between the in-situ and SWOT data sources assimilated in the FDA configuration.
}

\huy{The performance of SWDA, in terms of analyzed WLs at observing stations, is only comparable to that of the OL simulation. 
This stems from both the low revisit frequency of SWOT, which limits the regularity of observational updates needed for effective parameter correction in SWDA, and its incomplete spatial coverage of the domain by some SWOT passes (e.g. 42 and 113). This incompleteness shall be elaborated later in subsection~\ref{sssec:wse_real}.
It can be noted that the SWDA analysis tends to underestimate WLs at most observation stations, notably at TON, LR1, and LR0. This under-estimation primarily results from an overly low inflow discharge being introduced into the hydraulic domain, as previously illustrated in the bottom panel of Figure~\ref{fig:control_2024}.}




Figure~\ref{fig:H_tot} showed that both IDA and FDA experiments exhibit high accuracy in simulated WLs across the hydraulic domain---including at all seven gauge stations, with slightly higher $\mathrm{RMSE}$s at stations used only for validation (e.g., MD1 and LR1). On the other hand, the SWDA experiment yields a noticeable improvement over the OL configuration only at the MD0 station. This localized enhancement results from the adjustment of the associated friction zone, where $K_{s_3}$ by SWDA (red line) was reduced to approximately 28~\Ksunit~on Figure~\ref{fig:control_2024}.


\subsubsection{Impact on water surface elevation profile}
\label{sssec:wse_real}

\huy{The key strength of SWOT lies in its ability to provide spatially continuous WSE profiles along the river for each satellite pass.
Figure~\ref{fig:comparSWOT0}, Figure~\ref{fig:comparSWOT1}, and Figure~\ref{fig:comparSWOT2} present the SWOT node observations alongside equivalent WSEs simulated by the DA experiments. It is important to note that, unlike in the OSSE framework, portions of the river between TON and MD0, as well as between TON and LMA, are not covered in Figure~\ref{fig:comparSWOT0} and Figure~\ref{fig:comparSWOT1}, respectively, due to these sections falling outside the swath coverage of passes 42 and 113.}

\huy{Figure~\ref{fig:comparSWOT0} shows that the WSEs along the river's downstream part are over-estimated by the OL (orange line) and IDA (green line), in the river segments governed by $K_{s_4}$ and $K_{s_5}$. 
In contrast, SWDA (red line) and FDA (violet line) maintain close to the SWOT observations. Yet, to achieve this consistency, SWDA compensated by significantly reducing the inflow discharge (adjusting $\mu$ to 0.8), while also increasing $K_{s_5}$ to 75~\Ksunit. This behavior indicates a trade-off in SWDA, where performance at gauge stations (Table~\ref{tab:RMSE_real}) is sacrificed in favor of improved consistency with SWOT observations along the river profile. Another similar situation occurs during pass 42 on 2024-03-08, where SWDA reduces the inflow correction factor $\mu$ to 0.9 in order to better align with the partially observed WSE profile from SWOT  nodes.}
\huy{On the other hand, by leveraging both in-situ and SWOT data, the FDA experiment maintains strong performance across the metrics in the observation space, without the need to compromise between accuracy at gauge stations and alignment with SWOT-derived profiles.}

The $\mathrm{RMSE}$s computed against SWOT WSEs at river nodes are summarized by Table~\ref{tab:rmse_SWOT}.
\huy{
It was shown that while IDA reduces errors at the gauge locations, it still produces over-estimations of the WSE profile across large portions of the domain. In the downstream part, this discrepancy is likely related to uncertainties in the representation of the main channel bathymetry. As previously illustrated in Figure~\ref{fig:profile}, the actual up-to-date riverbed appears to lie below the elevation defined in the TELEMAC-2D model. 
In addition, the lack of gauge station at the transition between $K_{s_4}$ and $K_{s_5}$ zones can also lead to over-estimated WSEs, as both La Réole stations (LR0 and LR1) are positioned near the downstream boundary of the $K_{s_5}$ zone.   
In contrast, the assimilation of all SWOT river nodes enables a more accurate estimation of the friction coefficients required to reproduce the observed WSE profiles. While bathymetric imperfections persist, they are partially compensated by elevated Strickler values, in some case exceeding 70~\Ksunit. This adjustment reflects the well-known equifinality between friction and bathymetry in hydraulic modeling, as previously discussed in \cite{Garambois2015}.
}


\huy{
The added value of SWOT observations becomes evident when compared to in-situ data, as they enable correction of the entire longitudinal water surface profile. In the SWDA experiment (red lines), a minor discrepancy persists at the downstream end, as only $K_{s_5}$ and $\mu$ can be modified to be consistent with the downstream rating curve.
Water surface profiles from the FDA experiment (violet lines) are shown closely mirror those from SWDA, as also reflected by the $\mathrm{RMSE}$s in Table~\ref{tab:rmse_SWOT}. However, such strong performances depend on the availability and distribution of SWOT observations along the river. While the SWOT revisit frequency remains a limitation, the results presented in Figure~\ref{fig:H_tot} and Figure~\ref{fig:comparSWOTS6} (FDA in violet) clearly demonstrate that combining in-situ measurements with SWOT-derived observations significantly enhances the accuracy of water surface profiles. This consistency is also expected to be maintained under both low- and high-flow conditions, as demonstrated in the OSSE.}

\begin{figure*}[!ht]

\centering
\begin{subfigure}[b]{0.9\linewidth}
\includegraphics[trim=0 0.3cm 0 0, clip,width=\linewidth]{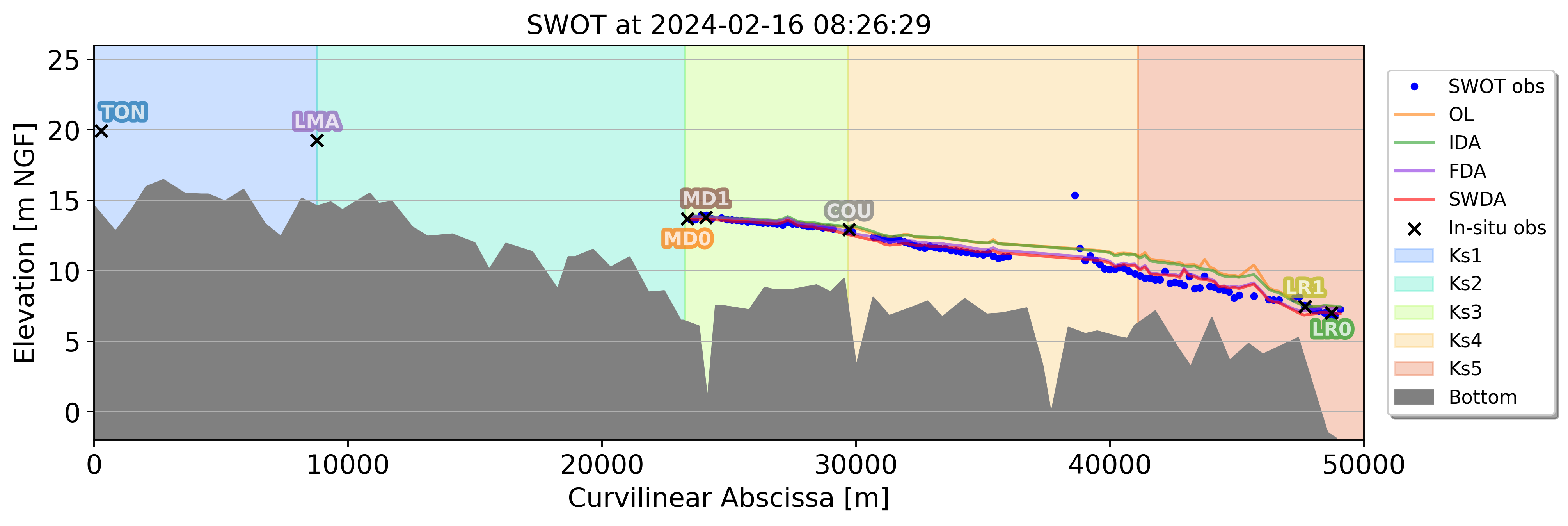}
\caption{SWOT observation on 2024-02-16 (pass 42)}\label{fig:comparSWOT0}
\end{subfigure}

\centering
\begin{subfigure}[b]{0.9\linewidth}
\includegraphics[trim=0 0.3cm 0 0, clip,width=\linewidth]{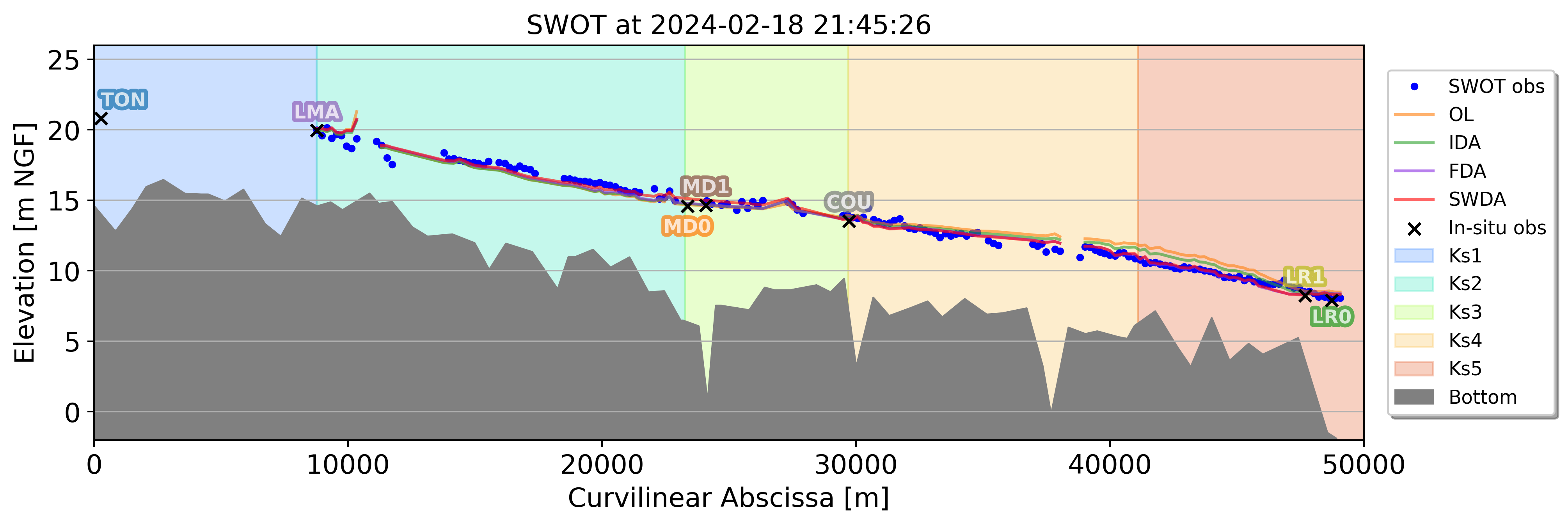}
\caption{SWOT observation on 2024-02-18 (pass 113)}\label{fig:comparSWOT1}
\end{subfigure}

\begin{subfigure}[b]{0.9\linewidth}
\includegraphics[trim=0 0.3cm 0 0, clip,width=\linewidth]{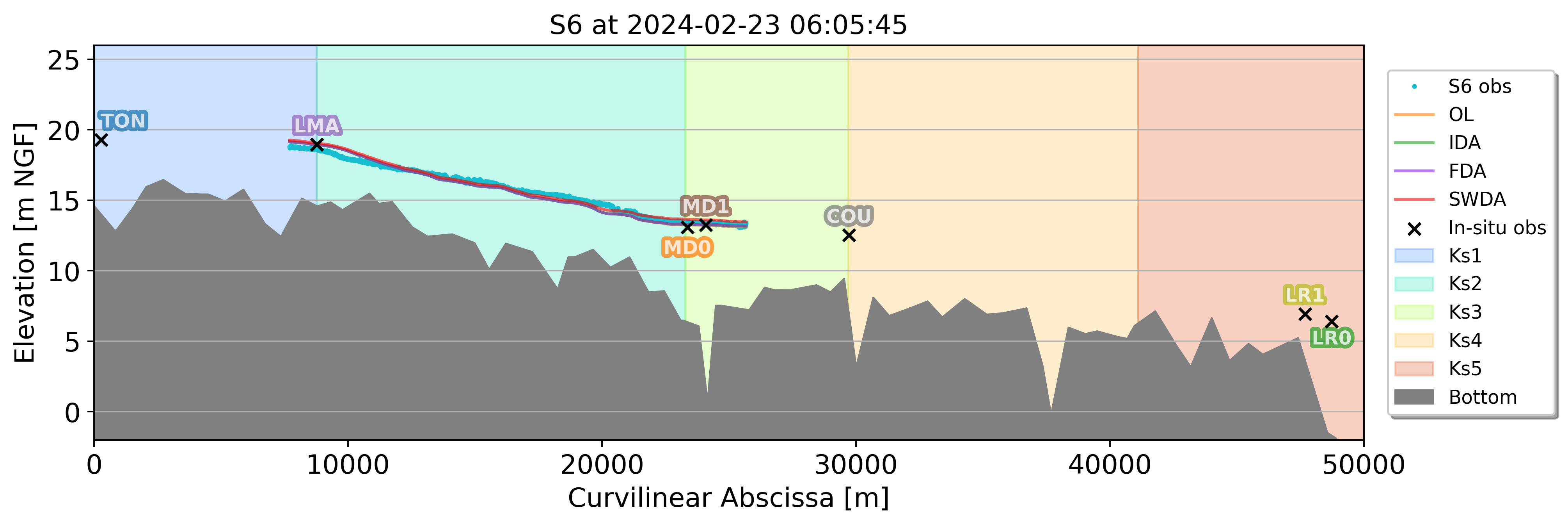}
\caption{Sentinel-6 observation on 2024-02-23}\label{fig:S6_1}
\end{subfigure}
\end{figure*}

\begin{figure*}[!ht]
\centering
\ContinuedFloat
\begin{subfigure}[b]{0.9\linewidth}
\includegraphics[trim=0 0.3cm 0 0, clip,width=\linewidth]{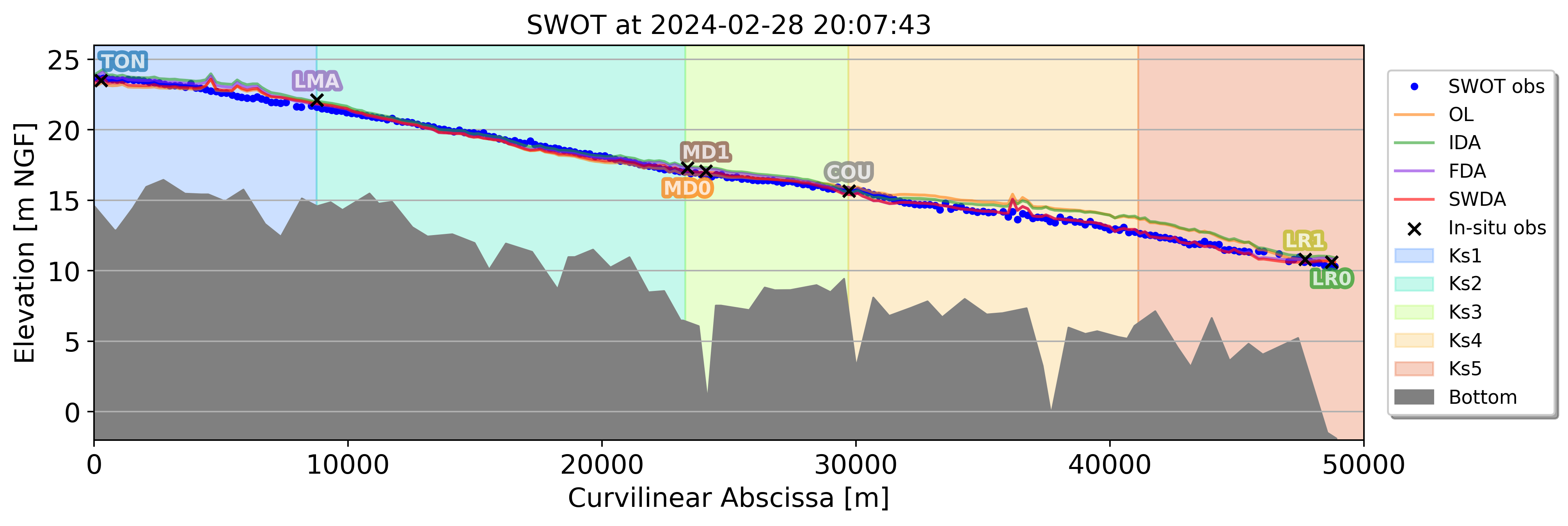}
\caption{SWOT observation on 2024-02-28 (pass 391)}\label{fig:comparSWOT2}
\end{subfigure}

\begin{subfigure}[b]{0.9\linewidth}
\includegraphics[trim=0 0.3cm 0 0, clip,width=\linewidth]{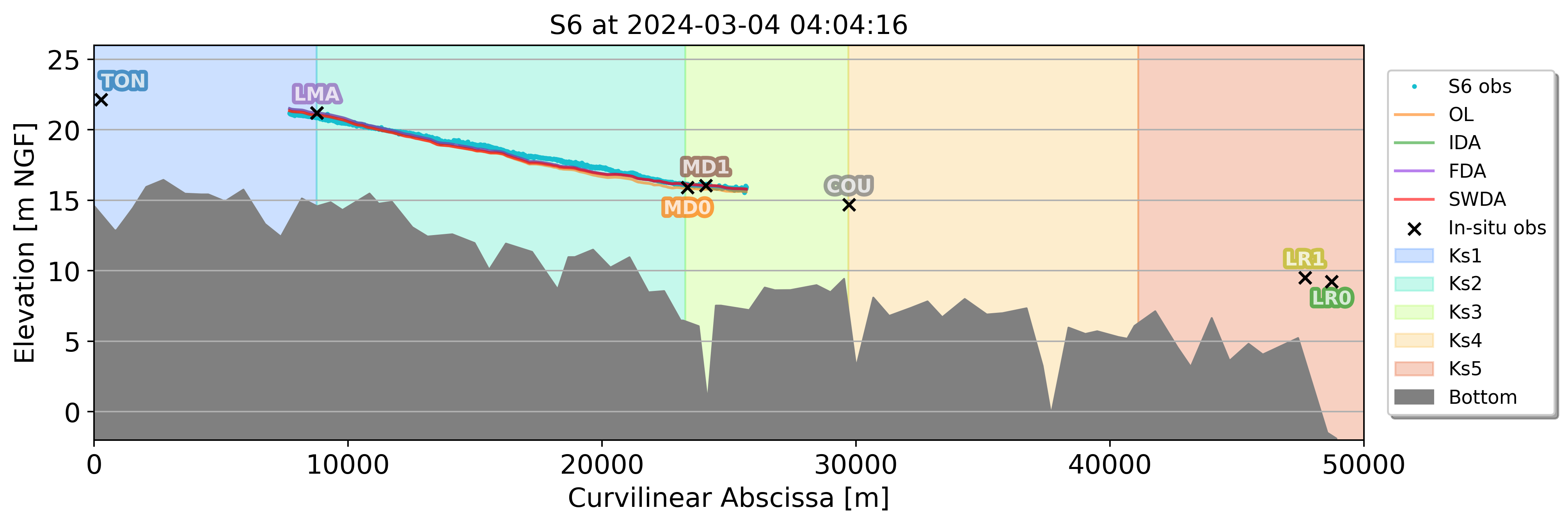}
\caption{Sentinel-6 observation on 2024-04-03}\label{fig:S6_2}
\end{subfigure}

\caption{WSE profiles between observed SWOT nodes (blue dots) or S6 FFSAR measurements (cyan dots) and OL (orange), IDA (green), SWDA (red) and FDA (violet). Circles indicate in-situ WSEs measured at gauge stations along the river reach.}
\label{fig:comparSWOTS6}
\end{figure*}

\huy{To support the assessment of SWOT-derived WSEs and the performance of the DA framework, independent validations were also conducted using Sentinel-6 altimetry data (processed with FFSAR) which observe the middle part of the domain between LMA and MD1. 
As such,  Figure~\ref{fig:S6_1} and Figure~\ref{fig:S6_2} present the comparison between Sentinel-6 WSE profiles (in cyan) along the river centerline, on 2024-02-23 and 2024-03-04, and the WSE profiles from the DA analyses.
The $\mathrm{RMSE}$s computed against all Sentinel-6 WSE profiles along the river centerline for all experiments are summarized by Table~\ref{tab:rmse_S6}.
Overall, a moderate agreement is observed between the Sentinel-6 WSE measurements and the simulated WSEs along the river centerline from the OL, with discrepancies reaching up to 34.7~cm. Similar discrepancies are noted in the IDA for several dates, although notably lower RMSE values, ranging between 20 and 23~cm, are achieved during higher-flow conditions on 2024-03-04 and 2024-03-14. 
Some pronounced deviations in the SWDA results are evident on 2024-02-23 and 2024-03-14, following the assimilation of SWOT observations from pass 113 on 2024-02-18 and 2024-03-10, respectively. These discrepancies likely arise from the missing SWOT coverage between the TON and LMA stations (also sparse around LMA), as illustrated in Figure~\ref{fig:comparSWOT1}, resulting in assimilation predominantly guided by SWOT nodes near the downstream station. Moreover, since only a single friction coefficient ($K_{s_2}$) governs this river segment with respect to the Sentinel-6 WSE profile, the lack of comprehensive SWOT data along the full reach restricts SWDA's capacity to effectively reduce errors across the entire stretch between LMA and MD1.}


\huy{In summary, the comparison between IDA and SWDA, or in other words between in-situ and RS observations, highlights the complementary strengths of each data source. In-situ measurements offer reliable WL data with high temporal sampling at discrete locations, yet they are limited in their ability to capture the continuous water surface profile between stations. While this limitation could be addressed by increasing the density of gauge stations, the associated costs are often prohibitive. Conversely, RS-derived WSE data (either from SWOT or Sentinel-6) provides large spatial coverage observations upon a catchment, albeit at lower temporal resolution. Results from the OSSE indicate that a denser revisit interval (ideally up to daily or sub-daily) would be sufficient to support high-quality reanalyses; however, such a satellite constellation is not currently operational. 
This combined strategy, as shown by FDA, enables robust temporal and spatial interpolation, effectively bridging the limitations of individual data sources. The findings presented here advocate that the proposed methodology---integrating model and multi-source data---can achieve the accuracy required for operational applications such as water balance estimation and flood risk management.}

    

\section{Conclusions}
\label{sec:conclusion}


This study highlights the merits of integrating satellite altimetry data---particularly from the recently launched SWOT mission---as a transformative resource for improving the representation of riverine flood hydrodynamics.
Here, we employed multi-source data assimilation within the framework of an EnKF integrated with the TELEMAC-2D hydrodynamic model over a reach of the Garonne River. Through various DA experiments conducted using both synthetic and real observational data, we have highlighted key findings that could inform future flood reanalysis and forecasting systems.  

The results from the OSSE confirm the ability of EnKF methods to improve model parameter estimation, especially when in-situ gauge data with high temporal resolution are assimilated (as in IDA experiment). In contrast, strategies relying solely on SWOT data (SWDA) were limited by the temporal resolution of satellite overpasses, leading to challenges in constraining hydraulic parameters during flood events.
The assimilation of both in-situ and SWOT-derived WSE data (FDA) proved to be the most effective strategy, offering robust parameter retrieval across a range of hydraulic conditions.

The study over the real 2024 event further validates the effectiveness of these methods. While the real data did not include overflowing into the floodplain, the results, in the control and observation spaces, show that the assimilation of SWOT in combination with in-situ WL measurements can accurately reproduce the river dynamics across different stations, offering promising prospects for future flood monitoring systems. 
Additionally, independent validations using Sentinel-6 measurements confirmed the model’s capability for temporal and spatial interpolation, even when assimilated solely SWOT-derived WSE data. This suggests that, with continued advancements in satellite altimetry, such models could be used effectively for real-time flood risk management and decision-making in ungauged areas.

This study also paves the way for future research exploring the integration of additional data sources, such as high-resolution flood extent observations or real-time sensor networks, to further refine flood forecasting models. As satellite altimetry technology continues to evolve, e.g. new hydrology-dedicated missions like SMASH, the prospects for integrating such data into operational flood management systems become even more promising, offering a pathway to more accurate and timely flood predictions.

\section*{Acknowledgments}
This work was partially funded by the Centre National d'Études Spatiales (CNES) and the Centre Européen de Recherche et de Formation Avancée en Calcul Scientifique (CERFACS) as part of the Space for Climate Observatory (SCO). The authors express their gratitude to Electricité de France (EDF) for supplying the TELEMAC-2D model for the Garonne Marmandaise catchment, as well as to SCHAPI, SPCs Garonne-Tarn-Lot, and Gironde-Adour-Dordogne for providing in-situ data. Additionally, they thank Vortex-io for providing the drone data, as well as J.-C. Poisson and V. Fouqueau (Vortex-io) for their assistance and advice regarding altimetric reference.



\bibliographystyle{IEEEtran}
\bibliography{refs}

\end{document}